\documentclass[a4paper]{article}

\usepackage[english]{babel}
\usepackage[utf8x]{inputenc}
\usepackage[T1]{fontenc}

\usepackage[a4paper,top=3.5cm,bottom=3.5cm,left=3.5cm,right=3.5cm,marginparwidth=1.75cm]{geometry}

\usepackage{amsmath}
\usepackage{graphicx}

\usepackage{pgf,tikz,colortbl,url,centernot}
\usepackage{amsthm}
\usepackage{natbib}
\usepackage{xcolor}

\newcommand{\ci}{\mathrel{\text{\scalebox{1.07}{$\perp\mkern-10mu\perp$}}}}

\usetikzlibrary{positioning}
\tikzstyle{lv}=[shape=circle,draw=black]
\tikzstyle{district}=[fill=black!20,shape=circle]

\title{\bf Graphical models for mediation analysis}

\author{
  Johan Steen\\
  \textit{\small Department of Intensive Care Medicine, Ghent University Hospital}
  \and
  Stijn Vansteelandt\\
  \textit{\small Department of Applied Mathematics, Computer Science and Statistics, Ghent University}\\
\textit{\small Centre for Statistical Methodology, London School of Hygiene and Tropical Medicine}
}

\begin{document}
\maketitle

\begin{abstract}
Mediation analysis seeks to infer how much of the effect of an exposure on an outcome can be attributed to specific pathways via intermediate variables or mediators. This requires identification of so-called path-specific effects. These express how a change in exposure affects those intermediate variables (along certain pathways), and how the resulting changes in those variables in turn affect the outcome (along subsequent pathways). However, unlike identification of total effects, adjustment for confounding is insufficient for identification of path-specific effects because their magnitude is also determined  by the extent to which individuals who experience large exposure effects on the mediator, tend to experience relatively small or large mediator effects on the outcome. This chapter therefore provides an accessible review of identification strategies under general nonparametric structural equation models (with possibly unmeasured variables), which rule out certain such dependencies. In particular, it is shown which path-specific effects can be identified under such models, and how this can be done.
\end{abstract}

\section{Introduction}

In many applications across a wide range of scientific disciplines, scholars aim to understand the mechanisms behind established cause-effect relationships, as witnessed by the widespread usage of mediation analyses.
Such understanding may not only be of pure scientific or etiologic interest, but may also aid policymakers in making informed decisions about public health interventions or reforms.

The Job Search Intervention Study (JOBS II), for instance, was designed to assess the effectiveness of a job training intervention to facilitate re-employment and reduce depressive symptoms in unemployed job seekers \citep{Vinokur1995a, Vinokur1997}.
1,249 randomly assigned job seekers were invited to participate in several sessions of job search skills workshops (the treatment group), whereas the remaining 552 unemployed workers received a booklet with job search tips (the control group).
Vinokur and Schul \cite{Vinokur1997} hypothesized that the treatment group would benefit from the workshops, assuming workshop attendance improves one's sense of self-efficacy and increases chances of getting re-employed, which, in turn, lead to a reduction in depressive symptoms.
Researchers thus believed an enhanced sense of mastery and re-employment to be active ingredients of the intervention's beneficial effect on mental health.

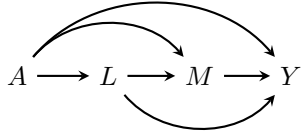
\begin{figure}[t!]
	\centering
	\begin{tikzpicture}[>=stealth,scale=0.8,thick]
		\node[] (a) at (0, 0)   {$A$};
		\node[] (l) at (1.5, 0)   {$L$};
		\node[] (m) at (3, 0)   {$M$};
		\node[] (y) at (4.5, 0)   {$Y$};
		\path[->] (a) edge node [above left] {} (l);
		\path[->, bend left=50] (a) edge node [above left] {} (m);
		\path[->, bend left=50] (a) edge node [above right] {} (y);
		\path[->] (l) edge node [above right] {} (m);
		\path[->, bend left=-50] (l) edge node [above right] {} (y);
		\path[->] (m) edge node [above right] {} (y);
	\end{tikzpicture}
	\caption{The treatment effect conceived as a combination of the effects along multiple causal chains. \label{pathways}}
\end{figure}

More generally, interventions or exposures essentially always realize their effects via a combination of causal chains or mechanisms.
Mediation analysis seeks to unravel and to quantify specific bundles of these pathways.
To fix ideas, consider the causal DAG in Figure~\ref{pathways}, which may represent hypothesized causal mechanisms underlying the effect of the job search intervention $A$ on mental health $Y$.
Suppose that workshop attendance increases participants' sense of self-efficacy $L$, which may, in turn, exert beneficial effects on mental health,
either by increasing chances of getting re-employed $M$ (along pathway $A\rightarrow L\rightarrow M\rightarrow Y$) or by other subsequent (unspecified) mechanisms (along pathway $A\rightarrow L\rightarrow Y$).
The intervention may also positively affect re-employment through other mechanisms before finally exerting its effect on mental health (along pathway $A\rightarrow M\rightarrow Y$),
or it may reduce depressive symptoms through none of the putative mediators (along pathway $A\rightarrow Y$).
Mediation analysis then aims to answer questions such as \textit{``How much of the intervention's effect on mental health is mediated by increased chances of re-employment?''}
It does so by disentangling the indirect effect that captures all pathways along which re-employment status $M$, the mediator of interest,
transmits the intervention effect ($A\rightarrow M\rightarrow Y$ and $A\rightarrow L\rightarrow M\rightarrow Y$) from the direct effect that captures all remaining pathways ($A\rightarrow Y$ and $A\rightarrow L\rightarrow Y$).
More generally, it aims to assess what effect the exposure realizes along one or multiple pathways.
We will informally refer to this as a \emph{path-specific effect} and give a precise definition later.

Bias-free estimation of path-specific effects crucially relies on certain structural assumptions and may often be compromised due to the subtle interplay between causal mechanisms.
In this chapter, we therefore aim to further elaborate on \emph{identifiability} of path-specific effects ---
that is, whether or not a certain set of causal assumptions suffices (or may even be deemed necessary) to identify such components from observed data.
In contrast to chapter 2, we wish to provide a more comprehensive overview of the literature on causal mediation analysis.\footnote{All references to other sections refer to the forthcoming \textit{Handbook of Graphical Models}, edited by Mathias
Drton, Steffen Lauritzen, Marloes Maathuis, and Martin Wainwright. It will contain this document as chapter 3 of Part IV.}
We first briefly review definitions of path-specific effects, in particular of natural direct and indirect effects \citep{Pearl2001,Greenland1992}, which are the standard targets of inference in causal mediation analysis.
Through various worked-out examples we next aim to develop intuition into non-parametric identification\footnote{For brevity, we will loosely use terms such as `identification', `identify', `recover from observed data' to refer to \emph{non-parametric} identification.}
of this class of path-specific effects and, in particular, the technical nature of certain assumptions on which mediation analysis relies.

\section{Definitions and notation}
To enable clear and formal definitions of the target causal estimands, let $A$ denote the exposure or treatment (e.g. workshop participation) and $Y$ the outcome of interest (e.g. presence of depressive symptoms).
Throughout, we will use counterfactual notation where, for instance, $Y(a)$ denotes the value of the outcome that would have been observed had $A$ (possibly contrary to the fact) been set to level $a$.
The (population-)\emph{average causal effect} can then be defined as $E\{Y(a)-Y(a')\}$, where $a$ and $a'$ correspond to meaningful levels of treatment.
This is essentially identical to the interventional contrast $E(Y\vert \text{do}(a)) - E(Y\vert \text{do}(a'))$ in terms of Pearl's do-operator.
For expositional simplicity, we will restrict our presentation to binary treatments (with $a=1$ and $a'=0$), although definitions and results extend to multicategorical or continuous treatments.
For instance, in our motivating example, $E\{Y(1)-Y(0)\}$ expresses the difference in prevalence of depressive symptoms if all unemployed workers were invited to participate in the job search skills workshop versus all received a booklet with job search tips.

\subsection{Natural direct and indirect effects}
Robins and Greenland \cite{Greenland1992} laid the foundations for effect decomposition by invoking \emph{nested counterfactuals} to conceptualize the intuitive notion of changing treatment assignment along specific pathways but not others.
For instance, the nested counterfactual $Y(a,M(a'))$ denotes the outcome that would have been observed had (possibly contrary to the fact) $A$ been set to level $a$ and $M$ to $M(a')$, the mediator value that had been observed had $A$ been set to $a'$.
Consequently, nested counterfactual expressions enable us to isolate and quantify part of the intervention effect that is transmitted through the mediator $M$ by leaving treatment unchanged at $A=1$, but changing the counterfactual intermediate outcome $M(1)$ to $M(0)$.
This then leads to the definition of the so-called \emph{natural indirect effect}
\begin{align*}
  E\{Y(1) - Y(1,M(0))\} = E\{Y(1,M(1)) - Y(1,M(0))\}
\end{align*}
Its complement, the \emph{natural direct effect}
\begin{align*}
   E\{Y(1) - Y(0)\} - E\{Y(1) - Y(1,M(0))\}&\\
   = E\{Y(1,M(0)) - Y(0)\} = E\{Y(1,M(0)) - Y(0,M(0))\}&
\end{align*}
captures the notion of blocking the intervention's effect on the mediator by keeping the latter fixed at whatever value it would have attained under no intervention.

In our motivating example, the natural direct effect expresses by how much the prevalence of depressive symptoms would change if all unemployed workers' treatment assignment status were to be changed, but their employment status were to be fixed to whatever status would be observed if they had originally received a booklet with job search tips.
In contrast, the natural indirect effect expresses the change in prevalence of depressive symptoms if all unemployed workers were to be invited to participate in the workshop, but their employment status were changed to whatever status would be observed if they had received a booklet with job search tips.

The main appeal of effect definitions that utilize nested counterfactuals, as opposed to equivalent formulations in the linear structural equation modeling tradition, is that they are model-free.
That is, they combine to produce the total effect, irrespective of the scale of interest or presence of interactions or nonlinearities, under the composition assumption that $Y(a,M(a)) = Y(a)$.
For instance, although the above effects are expressed in terms of mean (or risk) differences, the causal risk ratio of a binary outcome can similarly be expressed as the product of the natural direct effect risk ratio and the natural indirect effect risk ratio
\begin{align*}
\dfrac{E\{Y(1)\}}{E\{Y(0)\}} = \dfrac{E\{Y(1,M(0))\}}{E\{Y(0,M(0))\}} \dfrac{E\{Y(1,M(1))\}}{E\{Y(1,M(0))\}}.
\end{align*}

Consequently, mean nested counterfactuals can be parameterized using a class of marginal structural models \citep{Robins2000a} for mediation analyis, so-called \emph{natural effect models} \citep{Lange2013a,Lange2012,Loeys2013a,Steen2017a,Vansteelandt2012}, for instance
\begin{align}
E\{Y(a,M(a'))\} = g^{-1}(\beta_0 + \beta_1 a + \beta_2 a' + \beta_3 aa'),\label{nateffmod}
\end{align}
for all $a$ and $a'$ and where $g(\cdot)$ is a known link function.
If $g(\cdot)$ is chosen to be the identity link, $\beta_1$ captures the natural direct effect and $\beta_2+\beta_3$ captures the natural indirect effect on an additive scale.
Similarly, effects can be expressed on multiplicative scales, such as risk or odds ratios, by choosing $g(\cdot)$ to represent the log or logit link function.
Robins and Greenland \citep{Greenland1992} originally termed these parameters the \emph{pure direct effect} and \emph{total indirect effect}, respectively.
By differently apportioning the interaction term $\beta_3$, an alternative decomposition is obtained in terms of
the \emph{total direct effect} $E\{Y(1,M(1)) - Y(0,M(1))\}$, captured by $\beta_1+\beta_3$, and
the \emph{pure indirect effect} $E\{Y(0,M(1)) - Y(0,M(0))\}$, captured by $\beta_2$.
In accordance with VanderWeele \citep{VanderWeele2013a}, any of these two decompositions can thus be further refined, leading to the same unique three-way decomposition
into the pure direct effect $\beta_1$, the pure indirect effect $\beta_2$, and a mediated interactive effect $\beta_3$,
which can be interpreted to capture the extent to which direct and indirect pathways interact in their effect on the outcome.

Pearl \citep{Pearl2001} later adopted the same definitions but named these parameters \emph{natural} (rather than pure) direct and indirect effects
to emphasize that pure direct effects, as opposed to \emph{controlled direct effects} $E\{Y(1,m) - Y(0,m)\}$, allow for \emph{natural} variation in the mediator.
That is, natural direct effects reflect the effect of treatment upon fixing the mediator at values that would, for each individual, have \emph{naturally} occurred under no treatment,
rather than at some predetermined level $m$ (uniformly across the population).
In the remainder of this chapter, we will adopt Pearl's terminology of natural effects.

\subsection{Path-specific effects}
In graphical terms, the natural indirect effect quantifies the contribution along all pathways through which a single mediator transmits the treatment's effect on the outcome.
Its counterpart, the natural direct effect, quantifies the contribution along all remaining pathways from treatment to outcome.
Both of their counterfactual definitions refer to specific instances of nested counterfactuals of the form $Y(a,M(a'))$, with $a$ possibly different from $a'$.
Contributions along other predefined sets of directed paths $\pi$ from treatment $A$ to outcome $Y$ can similarly be defined in terms of contrasts of path-specific nested counterfactuals,
which we will denote $Y(\pi,a,a')$, in accordance with notation in the previous chapter.\footnote{For expositional simplicity, we will restrict settings to those with $A$ and $Y$ being singletons.}
As for natural effects, these $\pi$-specific counterfactuals represent two (possibly incompatible) hypothetical interventions which, for instance, set $A$ to $a$ for the purpose of all directed paths in $\pi$,
or to $a'$ for the purpose of directed paths not in $\pi$. For notational convenience, we denote $\overline\pi$ to be the set of directed pathways from $A$ to $Y$ not in $\pi$.

Suppose that, in our motivating example, interest lies in the effect of the job search intervention mediated by re-employment ($M$ in Figure~\ref{pathways}) but not by possible prior changes in perceived self-efficacy ($L$ in Figure~\ref{pathways}), as captured by $\pi = \{A\rightarrow M\rightarrow Y\}$.
This $\pi$-specific effect has been referred to as the \textit{partial} \citep{Huber2013} or \textit{semi-natural} \citep{Pearl2014} indirect effect with respect to $M$.
Its corresponding $\pi$-specific nested counterfactual $$Y(\{A\rightarrow M\rightarrow Y\},a,a') = Y(a',L(a'),M(a,L(a')))$$ can be obtained by recursive substitution, as discussed more formally in the previous chapter (Equation 1.5).
Just as counterfactuals of the form $Y(a,M(a'))$ give rise to definitions for natural effects, recursively nested counterfactuals of the above form enable us to define,
for instance, the \emph{pure} $\pi$-specific effect as
\begin{align*}
  E\{Y(0,L(0),M(1,L(0))) - Y(0)\}
\end{align*}
and the \emph{total} path-specific effect along pathways not in $\pi$ (or in $\overline\pi$) as
\begin{align*}
  E\{Y(1) - Y(0,L(0),M(1,L(0)))\}
\end{align*}
By symmetry and the composition assumption, these components again combine to produce the total effect of treatment.
The natural effect model $$E\{Y(a,L(a),M(a',L(a)))\} = \gamma_0 + \gamma_1 a + \gamma_2 a' + \gamma_3 aa',$$ for all $a$ and $a'$, maps these path-specific effects to $\gamma_1$ and $\gamma_2 + \gamma_3$, respectively.
Natural effect models that parameterize more fine-grained decompositions of the total causal effect into $k > 2$ path-specific effects (along $k-1$ ordered mediators) have been discussed in \cite{Steen2017a}.

\section{Cross-world quantities call for cross-world assumptions}
Despite the formal and intuitive appeal of path-specific effects, their non-parametric identification is subtle and a source of much controversy.
The reason is that the usual consistency assumptions alone --- for instance, that $M(a)=M$ when $A=a$ and that $Y(a,m)=Y$ when $A=a$ and $M=m$ --- do not suffice to link all counterfactuals to observed data.
In particular, nested counterfactual outcomes such as $Y(a,M(a'))$ are unobservable whenever $a\ne a'$.
Data, whether experimental or observational, thus never carry information about the distribution of these counterfactuals as they imply a union of two incompatible states $a$ and $a'$ that may only seem to coexist `across multiple worlds'.
Because of their `cross-world' nature, path-specific effects cannot in general be expressed in terms of interventional contrasts, which typically refer to ideal interventions in a single hypothetical world.
Mediation analyses based on natural or path-specific effects are thus bound to rely on assumptions that cannot be empirically verified or guaranteed by any study design \citep{Imai2013a,Greenland1992,Robins2010}.

To gradually develop intuition into non-parametric identification of natural and path-specific effects we will work through a number of simple but typical illustrative examples, spanning the next few sections of this chapter.
Unless stated otherwise, we shall assume throughout that treatment $A$ is randomized, in order to exclusively focus on assumptions characteristic to mediation analysis.
In this section, we particularly highlight that the distinct nature of nested counterfactuals calls for a type of assumption that cannot be verified empirically but that is, nonetheless, naturally encoded in so-called non-parametric structural equation models (NPSEMs).

\subsection{Imposing cross-world independence \label{firstex}}
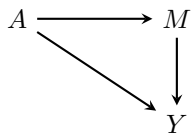
\begin{figure}[t!]
	\centering
	\begin{tikzpicture}[>=stealth,scale=0.7,thick]
		\node[] (a) at (0, 0)   {$A$};
		\node[] (m) at (3, 0)   {$M$};
		\node[] (y) at (3, -2)   {$Y$};
		\path[->] (a) edge node [above left] {} (m);
		\path[->] (a) edge node [above right] {} (y);
		\path[->] (m) edge node [above right] {} (y);
	\end{tikzpicture}
	\caption{A causal DAG representing a simple yet overly simplistic mediation setting. \label{exampleDAG0}}
\end{figure}

Identification of natural effects in the causal DAG $\mathcal{G}(\mathbf{V})$ with $\mathbf{V} = \{A,M,Y\}$, in Figure~\ref{exampleDAG0}, can be obtained if we recover the distribution $p(Y(a,M(a'))=y)$ of nested counterfactuals.
This requires summing (or integrating) the joint counterfactual distribution $p(Y(a,m)=y, M(a')=m)$ over $m$.
When $a\ne a'$, observed data carry no information about the dependence of $Y(a,m)$ on $M(a')$.
This articulates why natural effects cannot, in general, be identified from experimental data without further, untestable assumptions.
One such assumption is that of \emph{cross-world independence}
\begin{align}
	Y(a,m) \ci M(a').\tag{i}\label{cwass}
\end{align}
Under this assumption, we can factorize $p(Y(a,m)=y, M(a')=m)$ as a product of interventional distributions,
each of which is identified from observed data under the assumptions encoded in $\mathcal{G}(\mathbf{V})$, as follows
\begin{align*}
	p(Y(a,M(a'))=y) &= \sum_{m} p(Y(a,m)=y, M(a')=m)\\
	&= \sum_{m} p(Y(a,m)=y) p(M(a')=m) = \sum_{m} p(y\vert a,m) p(m\vert a'),
\end{align*}
where for arbitrary variables $V$ and $W$, $p(v\vert w)$ is shorthand notation for $p(V=v\vert W=w)$.\footnote{In this chapter, we will mostly use counterfactual notation instead of Pearl's do-notation,
especially when cross-world counterfactuals cannot be expressed using do-notation.
However, we will refer to counterfactual distributions as interventional distributions, whenever applicable.}
Pearl \citep{Pearl2001} claimed cross-world assumption~\eqref{cwass} to be key to `experimental' identification of natural effects.
With this, he indicated that, if interventional distributions $p(Y(a,m)=y)$ and $p(M(a')=m)$ were known from previous randomized interventions $\text{do}(a,m)$ and $\text{do}(a')$,
this assumption could be considered the missing link required to piece together these distributions in order to recover $p(Y(a,M(a'))=y)$.

\subsection{Cross-world independence and NPSEMs}
Subtleties surrounding cross-world assumptions such as~\eqref{cwass} have long been obscured to practitioners because of reliance on stringent parametric constraints or, more recently, on representations of causal DAGs as NPSEMs.
In fact, as discussed in more detail in the previous two chapters, NPSEMs impose restrictions on the joint distribution of all counterfactual outcomes,
including those that inhabit different worlds, that is, worlds under conflicting hypothetical interventions or treatment assignments such as $\text{do}(a)$ and $\text{do}(a')$.
As a result, cross-world independencies are naturally encoded by the set of (recursive) structural equations that defines a particular NPSEM.
For instance, the NPSEM representation of $\mathcal{G}(\mathbf{V})$ in Figure~\ref{exampleDAG0} is characterized by the following set of structural equations:
\begin{align*}
	A &:= f_A(\epsilon_A)				\\
	M &:= f_M(A, \epsilon_M)			\\
	Y &:= f_Y(A, M, \epsilon_Y)
\end{align*}
where $f_A$, $f_M$ and $f_Y$ are unknown deterministic functions and $\epsilon_A$, $\epsilon_M$ and $\epsilon_Y$ are mutually independent random error terms (representing unobserved background variables).
The assumed invariance of these equations endows them with a causal interpretation and permits us to deduce the counterfactual independencies they encode.
For example, under the interventions $\text{do}(a,m)$ and $\text{do}(a')$, the structural equations can respectively be written as
\begin{align*}
	A &:= a														& A &:= a'	\\
	M(a) &:= m													& M(a') &:= f_M(a', \epsilon_M)\\
	Y(a,m) &:= f_Y(a, m, \epsilon_Y)	& Y(a') &:= f_Y(a', M(a'), \epsilon_Y)
\end{align*}
Under this representation, the joint distribution of the one-step ahead counterfactuals
$$V(\mathbf{x}_{\text{pa}_{\mathcal{G}}(V)}) := f_{V}(\mathbf{x}_{\text{pa}_{\mathcal{G}}(V)}, \epsilon_V),$$
where $\text{pa}_{\mathcal{G}}(V)$ denotes the set of parents of $V$ in $\mathcal{G}(\mathbf{V})$
and $\mathbf{x}_{\text{pa}_{\mathcal{G}}(V)}$ the set of values to which these parents are set via the intervention $\text{do}(\mathbf{x})$,
is fully determined by the mutually independent error terms $\epsilon_{V}$.
It thus follows that all such one-step ahead counterfactuals are also mutually independent,
irrespective of the choice of hypothetical values $\mathbf{x}_{\text{pa}_{\mathcal{G}}(V)}$ to which we set the parents of $V$.
As a result, independence of the error terms $\epsilon_M \ci \epsilon_Y$ in the above structural equations
not only translates into $Y(a,m) \ci M(a)$ but also into cross-world independence (\ref{cwass}).
This may sound reassuring, but also signals the restrictiveness of NPSEMs,
as they inherently seem to encode independence assumptions that can never be verified from randomized interventions.

\subsection{Single world versus multiple worlds models}
Robins and Richardson \citep{Robins2010} extensively discuss these restrictions encoded by NPSEMs.
They moreover contrast the latter with another class of graphical causal models, Robins' \citep{Robins1986} \emph{Finest Fully Randomized Causally Interpretable Structured Tree Graph Model} (FFRCISTGM) representation of causal DAGs,
which only enforces restrictions that are (in principal) empirically verifiable.
Because this less restrictive class of models only imposes independence restrictions on sets of counterfactuals under a single set of (non-conflicting) interventions
such models have been referred to as `single world models', as opposed to NPSEMs which were termed `multiple worlds models'.
A more formal treatment of NPSEMs, FFRCISTGs and \emph{Single World Intervention Graphs} (SWIGs) \citep{Richardson2013},
which encode counterfactual independencies implied by a `single world model', is given in chapters 1, 2 and 5.

\subsection{Further outline}
Because NPSEMs naturally encode cross-world independence assumptions, they have provided a framework for the recent development of a fairly intuitive graphical rule
that governs whether and how nested `cross-world' counterfactual quantities relate to observed variables \citep{Avin2005}, even in the presence of unobserved or hidden variables \citep{Shpitser2013}.
In section~\ref{id2}, we demonstrate that specific cross-world independence assumptions can indeed be relatively easily interrogated from a (hidden variable) causal DAG interpreted as an NPSEM by this graphical rule.
As it turns out, this type of assumption forms the extra necessary layer on top of a set of assumptions that \emph{is} subject to experimental verification and serves to identify total treatment effects.
When combined with complete identification algorithms for total treatment effects \citep{Huang2006,Shpitser2006a,Tian2003}, the proposed graphical criterion therefore not only delineates sufficient, but also necessary conditions for identification of path-specific effects.

Essentially, when it comes to natural direct and indirect effects, this sound and complete criterion indicates that identification can be obtained under NPSEMs with the aid of two different types of auxiliary variables,
provided that no mediator-outcome confounders are themselves affected by treatment (and the total treatment effect is identifiable).
As the connection with earlier sufficient identification conditions for natural effects \citep{Pearl2001} seems to be somewhat missing from the literature,
we choose to review the main (two) graphical identification algorithms in chronological order (in sections~\ref{id1} and~\ref{id2}, respectively) and to revisit earlier assumptions in the light of this recent graphical criterion (in section~\ref{implcompl}).
In doing so, we point out that certain identification strategies have long been concealed because of the initial and exclusive focus on a single type of auxiliary variable
that recovers identification by establishing a conditional version of cross-world assumption~\eqref{cwass}.
Finally, in section~\ref{conccl}, we provide insights that may help to put a longstanding conceptual discussion regarding the very nature of mediation analysis into perspective.

\section{Identification 1.0 \label{id1}}
In this section, we further extend the simple causal DAG in Figure~\ref{exampleDAG0} to illustrate the logic and reasoning behind sufficient conditions for identification of natural effects.

\subsection{Unmeasured mediator-outcome confounding \label{unmeasuredconf}}
\begin{figure}[t!]
	\centering
	\begin{tikzpicture}[>=stealth,scale=0.75,thick]
    \node[] (title) at (-0.2, 1)   {(A)};
		\node[] (a) at (0, 0)   {$A$};
		\node[] (m) at (2, 0)   {$M$};
		\node[] (y) at (2, -2)   {$Y$};
		\node[lv] (u) at (4.5, 0)   {$U$};
		\path[->] (a) edge node [above left] {} (m);
		\path[->] (a) edge node [above right] {} (y);
		\path[->] (m) edge node [above right] {} (y);
		\path[->,dashed] (u) edge node [above right] {} (m);
		\path[->,dashed] (u) edge node [above right] {} (y);
	\end{tikzpicture}
	\hspace{0.3cm}
  \begin{tikzpicture}[>=stealth,scale=0.75,thick]
    \node[] (title) at (-0.2, 1)   {(B)};
    \node[] (a) at (0, 0)   {$A$};
    \node[] (m) at (2, 0)   {$M$};
    \node[] (y) at (2, -2)   {$Y$};
    \node[] (c) at (3.25, -1)   {$C$};
    \node[lv] (u) at (4.5, 0)   {$U$};
    \path[->] (a) edge node [above left] {} (m);
    \path[->] (a) edge node [above right] {} (y);
    \path[->] (m) edge node [above right] {} (y);
    \path[->] (c) edge node [above right] {} (y);
    \path[->,dashed] (u) edge node [above right] {} (m);
    \path[->,dashed] (u) edge node [above right] {} (c);
  \end{tikzpicture}
  \hspace{0.3cm}
  \begin{tikzpicture}[>=stealth,scale=0.75,thick]
    \node[] (title) at (-0.2, 1)   {(C)};
    \node[] (a) at (0, 0)   {$A$};
    \node[] (m) at (2, 0)   {$M$};
    \node[] (y) at (2, -2)   {$Y$};
    \node[] (c) at (3.25, 0)   {$C$};
    \node[lv] (u) at (4.5, 0)   {$U$};
    \path[->] (a) edge node [above left] {} (m);
    \path[->] (a) edge node [above right] {} (y);
    \path[->] (m) edge node [above right] {} (y);
    \path[->] (c) edge node [above right] {} (m);
    \path[->,dashed] (u) edge node [above right] {} (y);
    \path[->,dashed] (u) edge node [above right] {} (c);
  \end{tikzpicture}
	\caption{Causal DAGs that reflect more realistic mediation settings with unmeasured mediator-outcome confounding (A) along with two scenarios where a measured covariate $C$ may deconfound the mediator-outcome relation (B,C). \label{exampleDAG1}}
\end{figure}
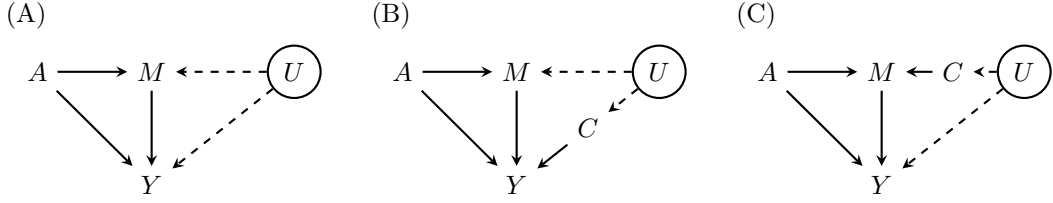

In most settings, the assumptions encoded by Figure~\ref{exampleDAG0} are unrealistic.
Indeed, even if treatment were randomized, as represented by the absence of back-door paths into $A$,
we cannot generally assume the absence of confounding of the mediator-outcome relation (other than by $A$) because typically $M$ is not randomized.
Nonetheless, independence of the error terms $\epsilon_M \ci \epsilon_Y$, as encoded in the NPSEM representation of Figure~\ref{exampleDAG0}, critically hinges on the assumption of \emph{no unmeasured mediator-outcome confounding}.
Because the latter assumption can be considered unlikely, the assumption of independent error terms is therefore almost guaranteed to be violated.
In this subsection, we will therefore relax assumptions by adding a hidden node $U$ that captures unmeasured confounding of the mediator-outcome relation (and induces dependence between their respective error terms when structural equations are expressed only in terms of observed variables $\mathbf{V}$), as in Figure~\ref{exampleDAG1}A.
More generally, we will represent unobserved variables $\mathbf{H}$ on hidden variable causal DAGs $\mathcal{G}(\textbf{V}\cup \textbf{H})$ by circled nodes.

By treatment randomization we have that $U \ci A$ such that, not surprisingly, the \textit{g-formula} \citep{Robins1986} yields
\begin{align}
p(Y(a)=y) = \sum_{u,m} p(y\vert a, m, u) p(m\vert a, u) p(u) = \sum_{m} p(y\vert a, m) p(m\vert a) = p(y\vert a).\label{eq1b}
\end{align}
Unfortunately, $U$ cannot similarly be integrated out `across worlds', which prevents us from identifying $p(Y(a,M(a'))=y)$,
even if conditional cross-world independence  $Y(a,m) \ci M(a') \vert U$ were to hold.
Indeed, we obtain
\begin{align}
p(Y(a,M(a'))=y) &= \sum_{u,m} p(Y(a,m)=y\vert u) p(M(a')=m\vert u) p(u)\nonumber\\
&= \sum_{u,m} p(y\vert a,m,u) p(m\vert a',u) p(u),\label{eq2}
\end{align}
an expression that cannot further be reduced to a functional of observed variables (such as Equation~\ref{eq1b}) because of the conflicting treatment assignments in its first two factors.

\subsection{Adjusting for mediator-outcome confounding \label{medformsection}}
Issues of non-identifiability of $p(Y(a,M(a'))=y)$ may, however, be remedied when one has available a measured set of prognostic covariates $\mathbf{C} \subseteq \mathbf{V}\setminus \{A,M,Y\}$
for mediator and/or outcome that renders the mediator-outcome relationship unconfounded given treatment assignment.
This can be understood because the availability of such a set $\mathbf{C}$, as, for instance, in the simplified example DAGs in Figures~\ref{exampleDAG1}B and~\ref{exampleDAG1}C where $\mathbf{C} = \{C\}$,
no longer necessitates stratifying on hidden variables such as $U$ to establish cross-world independence.

For example, in Figure~\ref{exampleDAG1}B, conditioning on $\mathbf{C}$ suffices, since the structural equations
\begin{align*}
	M(a') &:= f_M(a',U,\epsilon_M)\\
	Y(a,m) &:= f_Y(a,m,\mathbf{C},\epsilon_Y),
\end{align*}
indicate that cross-world independence holds within strata of $\mathbf{C}$, that is
\begin{align}
Y(a,m) \ci M(a')\vert \mathbf{C}.\tag{ii}\label{ccwass}
\end{align}
This then implies the same functional as expression~\eqref{eq2} but with unobserved $U$ replaced by the observed adjustment set $\mathbf{C}$
\begin{align}
p(Y(a,M(a'))=y) &= \sum_{\mathbf{c},m} p(Y(a,m)=y\vert \mathbf{c}) p(M(a')=m\vert \mathbf{c}) p(\mathbf{c})\label{intmedform}\\
&=\sum_{\mathbf{c},m} p(y\vert a,m,\mathbf{c}) p(m\vert a',\mathbf{c}) p(\mathbf{c}).\label{medform}
\end{align}
This functional is commonly referred to as Pearl's \cite{Pearl2001} \textit{mediation formula}.

To appreciate the importance of adjustment for prognostic factors $\mathbf{C}$, reconsider our motivating example.
Randomization of the intervention in itself did not suffice to eliminate potential confounding between re-employment $M$ and the outcome.
It is therefore essential to adjust for the pretreatment level of depression, a strong prognostic factor of the outcome of interest and most likely also related to re-employment.
Measurements on a range of other baseline covariates, including demographics, previous occupation and financial strain,
were also collected and adjusted for to strengthen the validity of cross-world assumption~\eqref{ccwass}.

\subsection{Treatment-induced mediator-outcome confounding \label{intermediateconf}}
The previous example may have led the reader to erroneously conclude that, given treatment randomization,
adjustment for a measured covariate set $\mathbf{C}$ that deconfounds the mediator-outcome relation within treatment arms
suffices to establish cross-world independence~\eqref{ccwass} under NPSEMs, thus enabling identification of $p(Y(a,M(a'))=y)$.
An important additional requirement is that no prognostic factor $L \in \mathbf{C}$ is affected by treatment.

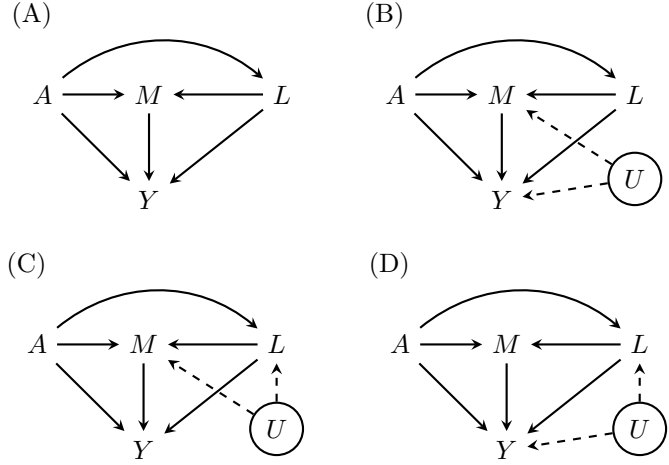
\begin{figure}[t!]
	\centering
	\begin{tikzpicture}[>=stealth,scale=0.7,thick]
		\node[] (title) at (-0.2, 1.5)   {(A)};
		\node[] (a) at (0, 0)   {$A$};
		\node[] (m) at (2, 0)   {$M$};
		\node[] (y) at (2, -2)   {$Y$};
		\node[] (l) at (4.5, 0)   {$L$};
		\path[->] (a) edge node [above left] {} (m);
		\path[->] (a) edge node [above right] {} (y);
		\path[->] (m) edge node [above right] {} (y);
		\path[->] (l) edge node [above right] {} (m);
		\path[->] (l) edge node [above right] {} (y);
		\path[->] (a) edge [bend left=40] node [above left] {} (l);
	\end{tikzpicture}
	\hspace{0.5cm}
	\begin{tikzpicture}[>=stealth,scale=0.7,thick]
		\node[] (title) at (-0.2, 1.5)   {(B)};
		\node[] (a) at (0, 0)   {$A$};
		\node[] (m) at (2, 0)   {$M$};
		\node[] (y) at (2, -2)   {$Y$};
		\node[] (l) at (4.5, 0)   {$L$};
		\node[lv] (u) at (4.5, -1.6)   {$U$};
		\path[->] (a) edge node [above left] {} (m);
		\path[->] (a) edge node [above right] {} (y);
		\path[->] (m) edge node [above right] {} (y);
		\path[->] (l) edge node [above right] {} (m);
		\path[->] (l) edge node [above right] {} (y);
		\path[->] (a) edge [bend left=40] node [above left] {} (l);
		\path[->,dashed] (u) edge node [above right] {} (m);
		\path[->,dashed] (u) edge node [above right] {} (y);
	\end{tikzpicture}
	\\ \vspace{0.3cm}
	\begin{tikzpicture}[>=stealth,scale=0.7,thick]
		\node[] (title) at (-0.2, 1.5)   {(C)};
		\node[] (a) at (0, 0)   {$A$};
		\node[] (m) at (2, 0)   {$M$};
		\node[] (y) at (2, -2)   {$Y$};
		\node[] (l) at (4.5, 0)   {$L$};
		\node[lv] (u) at (4.5, -1.6)   {$U$};
		\path[->] (a) edge node [above left] {} (m);
		\path[->] (a) edge node [above right] {} (y);
		\path[->] (m) edge node [above right] {} (y);
		\path[->] (l) edge node [above right] {} (m);
		\path[->] (l) edge node [above right] {} (y);
		\path[->] (a) edge [bend left=40] node [above left] {} (l);
		\path[->,dashed] (u) edge node [above right] {} (m);
		\path[->,dashed] (u) edge node [above right] {} (l);
	\end{tikzpicture}
	\hspace{0.5cm}
	\begin{tikzpicture}[>=stealth,scale=0.7,thick]
		\node[] (title) at (-0.2, 1.5)   {(D)};
		\node[] (a) at (0, 0)   {$A$};
		\node[] (m) at (2, 0)   {$M$};
		\node[] (y) at (2, -2)   {$Y$};
		\node[] (l) at (4.5, 0)   {$L$};
		\node[lv] (u) at (4.5, -1.6)   {$U$};
		\path[->] (a) edge node [above left] {} (m);
		\path[->] (a) edge node [above right] {} (y);
		\path[->] (m) edge node [above right] {} (y);
		\path[->] (l) edge node [above right] {} (m);
		\path[->] (l) edge node [above right] {} (y);
		\path[->] (a) edge [bend left=40] node [above left] {} (l);
		\path[->,dashed] (u) edge node [above right] {} (y);
		\path[->,dashed] (u) edge node [above right] {} (l);
	\end{tikzpicture}
	\caption{Causal DAGs that reflect mediation settings with treatment-induced confounding by $L$. \label{exampleDAG2}}
\end{figure}

Intuitively, if $L$ were a common cause of both $M$ and $Y$, as in the causal DAG $\mathcal{G}(\mathbf{V})$ with $\mathbf{V} = \{A,L,M,Y\}$, in Figure~\ref{exampleDAG2}A,
adjustment for $L$ (as in Equation~\ref{medform} with $\mathbf{C} = \{L\}$) would block the pathway $A\rightarrow L\rightarrow M\rightarrow Y$, which makes up part of the natural indirect effect of interest.
Lack of identification can formally be understood as follows.
According to the NPSEM associated with Figure~\ref{exampleDAG2}A, $Y(a,m) \ci M(a')$ holds conditional on $\{L(a)=l, L(a')=l'\}$ since
\begin{align*}
	M(a') &:= f_M(a',L(a'),\epsilon_M)\\
	Y(a,m) &:= f_Y(a,m,L(a),\epsilon_Y).
\end{align*}
Under the remaining assumptions encoded by this NPSEM, this allows us to express $p(Y(a,M(a'))=y)$ as
\begin{align*}
& \sum_{l,l',m} p(Y(a,m)=y\vert L(a)=l, L(a')=l')p(M(a')=m\vert L(a)=l, L(a')=l')\nonumber\\[-12pt] &\qquad \qquad \quad \times p(L(a)=l, L(a')=l')\nonumber\\
&= \sum_{l,l',m} p(y\vert a,l,m) p(m\vert a',l') p(L(a)=l, L(a')=l').
\end{align*}

As in the example of the previous section, this expression cannot further be reduced to a functional of the observed data
as it requires the joint cross-world counterfactual distribution $p(L(a)=l, L(a')=l')$.
Because this distribution again involves conflicting treatment assignments, strong untestable restrictions (beyond those encoded in NPSEMs) would be needed to enable identification.

In our motivating example, all available covariates were measured prior to randomization.
It may thus be safely assumed that none of them was affected by the intervention.
However, other mediators of the intervention's effect on mental health, such as an altered sense of self-efficacy, may well have affected re-employment
and thus manifest themselves as mediator-outcome confounders that are affected by the intervention.
In that case, cross-world independence~\eqref{ccwass} is likely violated.

\subsection{Pearl's graphical criteria for conditional cross-world independence\label{pearlgc}}
Pearl \cite{Pearl2001} devised two graphical criteria for assessing cross-world independence~\eqref{ccwass} under an NPSEM associated with a certain hidden variable causal DAG $\mathcal{G}(\mathbf{V}\cup\mathbf{H})$.
The logic for these criteria can be understood from the previous two examples in sections~\ref{unmeasuredconf} and~\ref{intermediateconf}.

The first criterion requires the availability of an adjustment set $\mathbf{C}$ that is sufficient, along with treatment $A$, to adjust for confounding of the mediator-outcome relation.
Such covariate set $\mathbf{C}$ should block all back-door paths between $M$ and $Y$ (except those traversing $A$) in the sense that
\begin{align}
  (Y \ci M\vert \mathbf{C})_{\mathcal{G}(\mathbf{V}\cup\mathbf{H})_{\underline{AM}}}.\tag{ii.a}\label{pc1}
\end{align}
That is, $\mathbf{C}$ d-separates $Y$ from $M$ in $\mathcal{G}(\mathbf{V}\cup\mathbf{H})_{\underline{AM}}$, the subgraph constructed from the original graph $\mathcal{G}(\mathbf{V}\cup\mathbf{H})$ by deleting all arrows emanating from $A$ and $M$.

The second criterion requires that
\begin{align}
  \text{no element of $\mathbf{C}$ is affected by treatment.}\tag{ii.b}\label{pc2}
\end{align}
We will henceforth refer to this criterion as `no treatment-induced confounding' or `no intermediate confounding'.

In the next two subsections, we review sufficient conditions for identifying natural direct and indirect effects from
i) experimental data from studies where treatment is randomized or from ii) observational data.
In doing so, we highlight that identification from purely observational data typically requires additional assumptions, which (in contrast to cross-world assumption~\eqref{ccwass}) are empirically falsifiable.
Following Pearl \citep{Pearl2014}, we compare different formulations of these additonal assumptions in terms of their identification power.

\subsection{Sufficient conditions to recover natural effects from experimental data}
Equation~\ref{intmedform} illustrates that cross-world independence~\eqref{ccwass} enables expressing the cross-world counterfactual distribution $p(Y(a,M(a'))=y)$
in terms of `single world' interventional distributions $p(M(a')=m\vert \mathbf{c})$ and $p(Y(a,m)=y\vert \mathbf{c})$.
It is easily demonstrated that these interventional distributions are identified under an NPSEM if treatment is randomized and cross-world independence~\eqref{ccwass} holds.
In other words, when combined with the ignorability condition that represents treatment randomization
\begin{align}
	\{Y(a,m),M(a),\mathbf{C}\} \ci A,\tag{iii}\label{treatrand}
\end{align}
conditional cross-world independence~\eqref{ccwass} enables identification of $p(Y(a,M(a'))=y)$ under an NPSEM from data obtained from a single randomized intervention $\text{do}(a)$ \citep{Imai2014}.
The latter implication could be considered an extension of Pearl's \citep{Pearl2001} `experimental' identification,
which formulates that marginal cross-world independence~\eqref{cwass} suffices to recover $p(Y(a,M(a'))=y)$ from two sequentially randomized interventions $\text{do}(a,m)$ and $\text{do}(a')$ (see section~\ref{firstex}; also see \citep{Imai2013a}).
This extension basically illustrates that, under a single randomized intervention $\text{do}(a)$, we need a measured set of baseline covariates $\mathbf{C}$
such that identification can be obtained under conditional cross-world independence~\eqref{ccwass}, without reliance on additional ignorability assumptions.

\subsection{Sufficient conditions to recover natural effects from observational data}
As opposed to randomized trials, assumption~\eqref{ccwass} is not sufficient for identifying natural effects from purely observational data.
This is because recovering interventional distributions $p(M(a')=m\vert \mathbf{c})$ and $p(Y(a,m)=y\vert \mathbf{c})$ from observational data
requires additional assumptions.

\subsubsection{The adjustment criterion for natural effects}
The availability of an adjustment set $\mathbf{C}$ that simultaneously satisfies assumption~\eqref{ccwass} and the following conditional ignorability assumption
\begin{align}
	\{Y(a,m), M(a)\} &\ci A\vert \mathbf{C}\tag{iv}\label{condtreatrand}
\end{align}
restores identifiability of $p(Y(a,M(a'))=y)$ under an NPSEM from observational data by the mediation formula (Equation~\ref{medform}) \citep{Imai2010a,Shpitser2011}.
Moreover, because, under NPSEMs, assumptions~\eqref{ccwass} and~\eqref{condtreatrand} are exchangeable (as a set) with the following set of conditional ignorability assumptions\footnote{Assumption~\eqref{adjform2} encodes cross-world independence, be it in a more subtle way.
That is, by the consistency assumption, it implies $Y(a,m)\ci M(a')\vert \{A=a',\mathbf{C}\}$, which is inherently cross-world counterfactual.} \citep{Shpitser2011}
\begin{align}
	M(a) &\ci A\vert \mathbf{C}\tag{v.a}\label{adjform1}\\
	Y(a,m) &\ci \{A,M\}\vert \mathbf{C},\tag{v.b}\label{adjform2}
\end{align}
it follows that the search for such a sufficient adjustment set $\mathbf{C}$ may as well be restricted to covariate sets
that simultaneously identify $p(M(a)=m)$ and $p(Y(a,m)=y)$ by the \emph{back-door} \citep{Pearl1995} or \emph{adjustment formula} \citep{Shpitser2010}.
This led Shpitser and VanderWeele \cite{Shpitser2011} to develop a complete graphical criterion for identification of $p(Y(a,M(a'))=y)$ by the mediation formula (under NPSEMs).
They instead termed this the \textit{adjustment formula for natural direct and indirect effects} as it generalizes the adjustment criterion for total effects \citep{Shpitser2010} to mediation settings.

Intuitively, this criterion can be thought of as aiming to establish both cross-world independence~\eqref{ccwass}
and conditions~\eqref{adjform1} and~\eqref{adjform2} solely by means of adjustment for a common measured covariate set $\mathbf{C}$.
First, it demands no unmeasured mediator-outcome confounding (as in Figure~\ref{exampleDAG1}A),
which, if not met, violates cross-world independence~\eqref{ccwass} and, moreover, hampers identification of $p(Y(a,m)=y)$ by the adjustment formula.
Second, it demands the absence of treatment-induced mediator-outcome confounders (such as $L$ in Figure~\ref{exampleDAG2}A),
because their presence both violates cross-world independence~\eqref{ccwass} and hinders the availability of a common set $\mathbf{C}$
that enables identification of both $p(M(a)=m)$ and $p(Y(a,m)=y)$ by the adjustment formula.
Crucially, establishing cross-world independence~\eqref{ccwass} and conditions~\eqref{adjform1} and~\eqref{adjform2} by means of this generalized adjustment criterion goes hand in hand.

\begin{figure}[t!]
	\centering
	\begin{tikzpicture}[>=stealth,scale=1,thick]
		\node[] (title) at (-0.2, 2.1)   {$\mathcal{G}(\mathbf{V}\cup \mathbf{H})$};
		\node[] (a) at (0, 0)   {$A$};
		\node[] (m) at (3, 0)   {$M$};
		\node[] (y) at (3, -2)   {$Y$};
		\node[] (c1) at (5, 0)   {$C_1$};
		\node[] (c2) at (3, 2)   {$C_2$};
		\node[] (c3) at (0, -2)   {$C_3$};
		\node[lv] (u1) at (1.95, 0.6)   {$U_1$};
		\node[lv] (u2) at (3.6, -0.56)   {$U_2$};
		\node[lv] (u3) at (4, 1)   {$U_3$};
		\path[->] (a) edge node [above left] {} (m);
		\path[->] (a) edge node [above right] {} (y);
		\path[->] (m) edge node [above right] {} (y);
		\path[->] (c1) edge node [above right] {} (m);
		\path[->] (c1) edge node [above right] {} (y);
		\path[->] (c2) edge node [above right] {} (a);
		\path[->] (c2) edge node [above right] {} (m);
		\path[->] (c3) edge node [above right] {} (a);
		\path[->] (c3) edge node [above right] {} (y);
		\path[->,dashed] (u1) edge node [above right] {} (c2);
		\path[->,dashed] (u1) edge node [above right] {} (c3);
		\path[->,dashed] (u2) edge node [above right] {} (c1);
		\path[->,dashed] (u2) edge node [above right] {} (c3);
		\path[->,dashed] (u3) edge node [above right] {} (c1);
		\path[->,dashed] (u3) edge node [above right] {} (c2);
	\end{tikzpicture}
	\hspace{10pt}
	\begin{tikzpicture}[>=stealth,scale=1,thick]
		\node[] (title) at (-0.2, 2.1)   {$\mathcal{G}(\mathbf{V}\cup \mathbf{H})_{\underline{AM}}$};
		\node[] (a) at (0, 0)   {$A$};
		\node[] (m) at (3, 0)   {$M$};
		\node[] (y) at (3, -2)   {$Y$};
		\node[] (c1) at (5, 0)   {$C_1$};
		\node[] (c2) at (3, 2)   {$C_2$};
		\node[] (c3) at (0, -2)   {$C_3$};
		\node[lv] (u1) at (1.95, 0.6)   {$U_1$};
		\node[lv] (u2) at (3.6, -0.56)   {$U_2$};
		\node[lv] (u3) at (4, 1)   {$U_3$};
		\path[->] (c1) edge node [above right] {} (m);
		\path[->] (c1) edge node [above right] {} (y);
		\path[->] (c2) edge node [above right] {} (a);
		\path[->] (c2) edge node [above right] {} (m);
		\path[->] (c3) edge node [above right] {} (a);
		\path[->] (c3) edge node [above right] {} (y);
		\path[->,dashed] (u1) edge node [above right] {} (c2);
		\path[->,dashed] (u1) edge node [above right] {} (c3);
		\path[->,dashed] (u2) edge node [above right] {} (c1);
		\path[->,dashed] (u2) edge node [above right] {} (c3);
		\path[->,dashed] (u3) edge node [above right] {} (c1);
		\path[->,dashed] (u3) edge node [above right] {} (c2);
	\end{tikzpicture}
	\caption{Hidden variable causal DAG $\mathcal{G}(\mathbf{V}\cup \mathbf{H})$ which permits identification of $p(Y(a,M(a'))=y)$ by the adjustment criterion (under its NPSEM representation).
	The subgraph $\mathcal{G}(\mathbf{V}\cup \mathbf{H})_{\underline{AM}}$ aids in selecting a candidate covariate set that satisfies cross-world independence~\eqref{ccwass} by Pearl's graphical criteria described in section~\ref{pearlgc}.
  \label{exampleDAGpiecemeal}}
\end{figure}
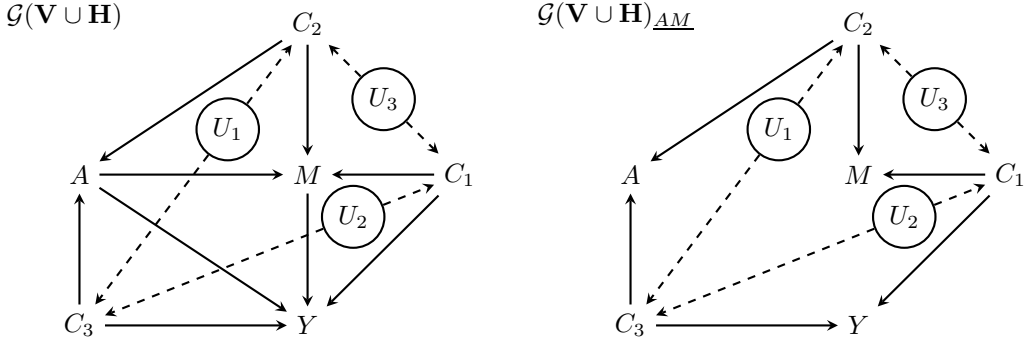

\paragraph{\bf A relatively simple example}
Consider the causal DAG $\mathcal{G}(\mathbf{V}\cup\mathbf{H})$ with observed variables $\mathbf{V} = \{C_1,C_2,C_3,A,M,Y\}$ and hidden variables $\mathbf{H} = \{U_1,U_2,U_3\}$
in Figure~\ref{exampleDAGpiecemeal} (adopted from \citep{Pearl2014}; Figure 3B).
The search for a candidate covariate set $\mathbf{C}$ that satisfies assumption~\eqref{ccwass} may be guided by graphical criteria~\eqref{pc1} and~\eqref{pc2}, as discussed in section~\ref{pearlgc}.
Because $\mathcal{G}(\mathbf{V}\cup\mathbf{H})$ includes no intermediate confounders, it suffices to search for a set of baseline covariates that d-separates $M$ from $Y$ in the subgraph $\mathcal{G}(\mathbf{V}\cup\mathbf{H})_{\underline{AM}}$.
These candidate adjustment sets include $\{C_1,C_2\}$, $\{C_1,C_3\}$ and $\{C_1,C_2,C_3\}$.
Assessing whether $p(Y(a,M(a'))=y)$ can be recovered from observed data by the mediation formula
now boils down to verifying whether both $p(M(a)=m)$ and $p(Y(a,m)=y)$ are identified by the adjustment formula upon adjustment for one of these candidate sets.
It turns out that only the set $\{C_1,C_2,C_3\}$ identifies both of these interventional distributions by the adjustment formula,
such that under the NPSEM representation of $\mathcal{G}(\mathbf{V}\cup\mathbf{H})$, we obtain
\begin{align*}
  p(Y(a,M(a'))=y) &= \sum_{c_1,c_2,c_3,m} p(y\vert a,m,c_1,c_2,c_3) p(m\vert a',c_1,c_2,c_3) p(c_1,c_2,c_3).
\end{align*}
In fact, under NPSEMs, any covariate set $\mathbf{C}$ that suffices to identify both $p(M(a)=m)$ and $p(Y(a,m)=y)$ by the adjustment formula,
will also satisfy cross-world independence~\eqref{ccwass}, such that the initial step can simply be skipped \citep{Shpitser2011}.

\subsubsection{Identification beyond the adjustment criterion \label{idbeyondadjcrit}}
A major appeal of identification via the adjustment criterion for natural effects is that it leads to a standard identifying functional.
This, in turn, allows for general modeling and estimation strategies.
However, as the following examples illustrate, it may unnecessarily increase modeling demands and limit the ability to identify $p(Y(a,M(a'))=y)$.

\paragraph{\bf A relatively simple example revisited}
By exploiting the following exclusion restrictions encoded in $\mathcal{G}(\mathbf{V}\cup\mathbf{H})$ in Figure~\ref{exampleDAGpiecemeal}
\begin{align*}
	M &\ci C_3\vert \{A,C_1,C_2\}\\
	Y &\ci C_2\vert \{A,M,C_1,C_3\},
\end{align*}
the identification result can be simplified as
\begin{align*}
  &\sum_{c_1,c_2,c_3,m} p(y\vert a,m,c_1,c_3) p(m\vert a',c_1,c_2) p(c_1,c_2,c_3),
\end{align*}
thereby reducing modeling demands (although see \citep{Imai2014} for a critical discussion on such exclusion restrictions).

\begin{figure}[t!]
 	\centering
 	 	\begin{tikzpicture}[scale=0.87,>=stealth,thick]
 	 	\node [anchor=west] at (-3.2, 2.5) {$\mathcal{G}(\mathbf{V}\cup \mathbf{H})$};
 		\node[] (a) at (0, 0)   {$A$};
 		\node[] (m) at (3, 0)   {$M$};
 		\node[] (y) at (3, -2)   {$Y$};
 		\node[] (w1) at (-2, 1.5)   {$C_1$};
 		\node[] (w2) at (-2, 0)   {$C_2$};
 		\node[] (w3) at (-2, -2)   {$C_3$};
 		\node[lv] (u1) at (-2, -1)   {$U_1$};
 		\node[lv] (u2) at (0.5, 0.75)   {$U_2$};
 		\node[lv] (u3) at (-1, -1)   {$U_3$};
 		\path[->] (a) edge node [above left] {} (m);
 		\path[->] (a) edge node [above right] {} (y);
 		\path[->] (m) edge node [above right] {} (y);
 		\path[->] (w1) edge [bend left] node [above right] {} (m);
 		\path[->] (w1) edge [bend right=45] node [above right] {} (w3);
 		\path[->] (w2) edge node [above right] {} (a);
 		\path[->] (w3) edge node [above right] {} (y);
 		\path[->,dashed] (u1) edge node [above right] {} (w2);
 		\path[->,dashed] (u1) edge node [above right] {} (w3);
 		\path[->,dashed] (u2) edge node [above right] {} (w2);
 		\path[->,dashed] (u2) edge node [above right] {} (m);
 		\path[->,dashed] (u3) edge node [above right] {} (a);
 		\path[->,dashed] (u3) edge node [above right] {} (w3);
 	\end{tikzpicture}
	\hspace{20pt}
    \begin{tikzpicture}[scale=0.87,>=stealth,thick]
    \node [anchor=west] at (-3.2, 2.5) {$\mathcal{G}(\mathbf{V})$};
    \node[] (a) at (0, 0)   {$A$};
    \node[] (m) at (3, 0)   {$M$};
    \node[] (y) at (3, -2)   {$Y$};
    \node[] (w1) at (-2, 1.5)   {$C_1$};
    \node[] (w2) at (-2, 0)   {$C_2$};
    \node[] (w3) at (-2, -2)   {$C_3$};
    \path[->] (a) edge node [above left] {} (m);
    \path[->] (a) edge node [above right] {} (y);
    \path[->] (m) edge node [above right] {} (y);
    \path[->] (w1) edge [bend left] node [above right] {} (m);
    \path[->] (w1) edge [bend right=45] node [above right] {} (w3);
    \path[->] (w2) edge node [above right] {} (a);
    \path[->] (w3) edge node [above right] {} (y);
    \path[<->,dashed] (a) edge node [above right] {} (w3);
    \path[<->,dashed] (w2) edge [bend left=25] node [above right] {} (m);
    \path[<->,dashed] (w2) edge node [above right] {} (w3);
    \end{tikzpicture}
	\\ \vspace{7pt}
	\begin{tikzpicture}[scale=0.87,>=stealth,thick]
		\node [anchor=west] at (-3.2, 2.5) {$\mathcal{G}(\mathbf{V})_{\mathbf{V}\setminus A}$};
		\node[] (m) at (3, 0)   {$M$};
		\node[] (y) at (3, -2)   {$Y$};
		\node[] (w1) at (-2, 1.5)   {$C_1$};
		\node[] (w2) at (-2, 0)   {$C_2$};
		\node[] (w3) at (-2, -2)   {$C_3$};
		\path[->] (m) edge node [above right] {} (y);
		\path[->] (w1) edge [bend left] node [above right] {} (m);
		\path[->] (w1) edge [bend right=45] node [above right] {} (w3);
		\path[->] (w3) edge node [above right] {} (y);
    \path[<->,dashed] (w2) edge [bend left=25] node [above right] {} (m);
    \path[<->,dashed] (w2) edge node [above right] {} (w3);
	\end{tikzpicture}
	\hspace{20pt}
	\begin{tikzpicture}[scale=0.87,>=stealth,thick]
  	\node [anchor=west] at (-3.2, 2.5) {$\mathcal{G}(\mathbf{V})_\mathbf{Y^*}$};
  	\node[] (m) at (3, 0)   {$M$};
  	\node[] (y) at (3, -2)   {$Y$};
  	\node[] (w1) at (-2, 1.5)   {$C_1$};
  	\node[] (w3) at (-2, -2)   {$C_3$};
  	\path[->] (m) edge node [above right] {} (y);
  	\path[->] (w1) edge [bend left] node [above right] {} (m);
  	\path[->] (w1) edge [bend right=45] node [above right] {} (w3);
  	\path[->] (w3) edge node [above right] {} (y);
	\end{tikzpicture}
 	\caption{A somewhat more involved hidden variable causal DAG $\mathcal{G}(\mathbf{V}\cup \mathbf{H})$,
  its latent projection ADMG $\mathcal{G}(\mathbf{V})$ and subgraphs $\mathcal{G}(\mathbf{V})_{\mathbf{V}\setminus A}$ and $\mathcal{G}(\mathbf{V})_\mathbf{Y^*}$.
  \label{Pearl2014fig5fsingle}}
\end{figure}
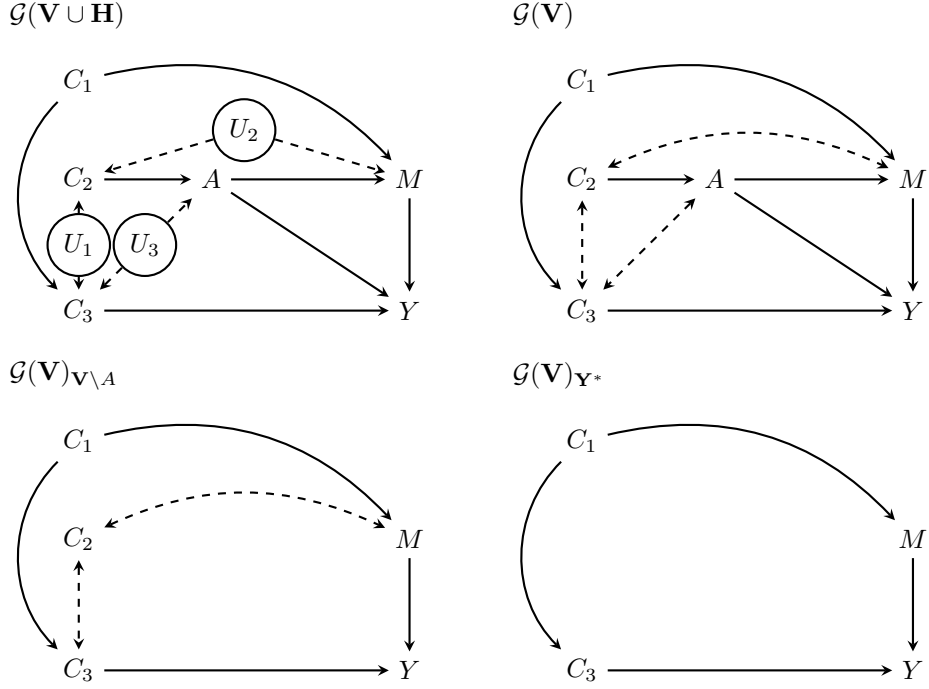

\paragraph{\bf A somewhat more involved example}
Consider next the causal DAG $\mathcal{G}(\mathbf{V}\cup\mathbf{H})$ with observed variables $\mathbf{V} = \{C_1,C_2,C_3,A,M,Y\}$ and hidden variables $\mathbf{H} = \{U_1,U_2,U_3\}$ in Figure~\ref{Pearl2014fig5fsingle} (adopted from \citep{Pearl2014}; Figure 5F).
To identify $p(M(a)=m)$ we must, in any case, adjust for $C_2$. Since $C_3$ is a collider, adjusting for it opens spurious pathways
that cannot be blocked by additionally adjusting for $C_1$, leaving $\{C_2\}$ and $\{C_1,C_2\}$ as the only viable adjustment sets for identifying $p(M(a)=m)$ by the adjustment formula.
However, identification of $p(Y(a,m)=y)$ by the adjustment formula requires that $C_3$ is included in the adjustment set,
because the back-door path from $A$ to $Y$ via $U_3$ can only be blocked by $C_3$.
As a result, $p(Y(a,M(a'))=y)$ cannot be identified by the adjustment criterion for natural direct and indirect effects.

Nonetheless, identification can be obtained by resorting to an alternative identification strategy.
Such strategy may consist of first listing all sufficient adjustment sets that identify $p(M(a)=m)$ and all sufficient adjustment sets that identify $p(Y(a,m)=y)$.
Progress can then be made if a subset of the intersection of any two of these respective candidate adjustment sets satisfies assumption~\eqref{ccwass}.
For instance, $p(M(a)=m)$ is identified by adjusting for $\{C_1,C_2\}$, whereas $p(Y(a,m)=y)$ is identified by adjusting for $\{C_1,C_3\}$.
Moreover, $\{C_1\}$, the intersection of these separate adjustment sets, satisfies cross-world independence~\eqref{ccwass}.
Relying on the conditional independence $Y \ci C_1 \vert \{A,M,C_3\}$, $p(Y(a,M(a'))=y)$ may then be identified from observed data by
\begin{align} \label{invexnatural}
  & \sum_{c_1,m} p(Y(a,m)=y\vert c_1) p(M(a')=m\vert c_1) p(c_1)\nonumber\\[-5pt]
	&= \sum_{c_1,m} \left(\sum_{c_3} p(y\vert a,m,c_1,c_3)p(c_3\vert c_1)\right) \left(\sum_{c_2} p(m\vert a',c_1,c_2) p(c_2)\right) p(c_1)\nonumber\\
	&= \sum_{c_1,c_2,c_3,m} p(y\vert a,m,c_3) p(m\vert a',c_1,c_2) p(c_1) p(c_2) p(c_3\vert c_1).
\end{align}

The above examples demonstrate that $p(M(a')=m\vert \mathbf{c})$ and $p(Y(a,m)=y\vert \mathbf{c})$ in Equation~\ref{intmedform}
can be identified under a much wider range of scenarios than those that lead to identification by the adjustment criterion for natural effects \citep{Pearl2014}.
That is, identification of $p(Y(a,M(a'))=y)$ can be obtained if, for any candidate covariate set $\mathbf{C}$ that satisfies~\eqref{ccwass},
$p(M(a')=m\vert \mathbf{c})$ and $p(Y(a,m)=y\vert \mathbf{c})$ are identified by Shpitser's complete \textbf{IDC} algorithm for conditional treatment effects \citep{Shpitser2006} (see section 1.3.6 of the previous chapter).
This resonates Pearl's original formulations \citep{Pearl2001} which state that, to recover `non-experimental' identification,
assumption~\eqref{ccwass} needs to be complemented with the following two assumptions
\begin{align}
  p(M(a)=m\vert \mathbf{c}) \text{ is identifiable by some means, and}&\tag{vi.a}\label{mac}\\
  p(Y(a,m)=y\vert \mathbf{c}) \text{ is identifiable by some means.}&\tag{vi.b}\label{yamc}
\end{align}

One way to increase identification power is by resorting to the `divide and conquer' strategy described in the previous example,
which Pearl \citep{Pearl2014} referred to as \emph{piecemeal deconfounding}.
However, in certain settings, this strategy may still be overly restrictive and identification may then, instead,
sometimes be recovered by exploiting so-called \emph{mediating instruments} \citep{Pearl2014}.
That is, if specific instruments can be found that fully mediate certain crucial but confounded paths (that cannot be deconfounded by observed covariates),
further progress can be made by local application of the front-door formula.
Specific examples are given in \cite{Pearl2014}.

\section{Identification 2.0 \label{id2}}
Widening the scope to also include mediating instruments in our `identification toolbox' still does not enable us
to fully characterize all possible settings that enable non-parametric identification of $p(Y(a,M(a')))$ under NPSEMs.
In other words, while the assumption set~\eqref{ccwass}-\eqref{mac}-\eqref{yamc} may be sufficient for recovering natural effects
from observational data, it is not necessary and, consequently, not \emph{complete} for identification.
This lack of completeness can be demonstrated using a simple illustrating example.

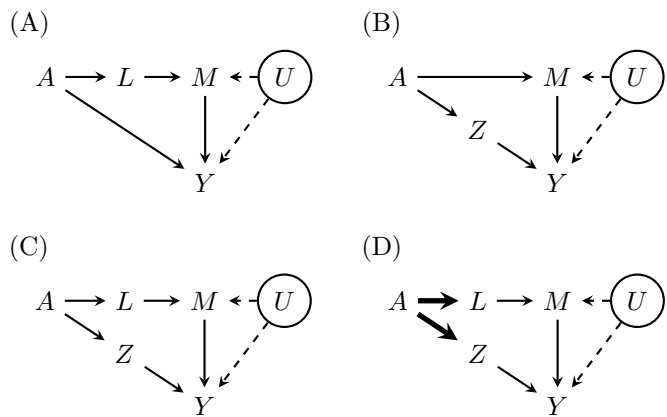
\begin{figure}[t!]
	\centering
	\begin{tikzpicture}[>=stealth,scale=0.7,thick]
    \node[] (title) at (-0.3, 1)   {(A)};
    \node[] (a) at (0, 0)   {$A$};
		\node[] (m) at (3, 0)   {$M$};
		\node[] (y) at (3, -2)   {$Y$};
		\node[] (l) at (1.5, 0)   {$L$};
    \node[lv] (u) at (4.5, 0)   {$U$};
		\path[->] (a) edge node [above right] {} (y);
		\path[->] (m) edge node [above right] {} (y);
		\path[->] (l) edge node [above right] {} (m);
		\path[->] (a) edge node [above left] {} (l);
    \path[->,dashed] (u) edge node [above right] {} (m);
    \path[->,dashed] (u) edge node [above right] {} (y);
	\end{tikzpicture}
	\hspace{0.3cm}
  \begin{tikzpicture}[>=stealth,scale=0.7,thick]
    \node[] (title) at (-0.3, 1)   {(B)};
    \node[] (a) at (0, 0)   {$A$};
    \node[] (m) at (3, 0)   {$M$};
    \node[] (y) at (3, -2)   {$Y$};
    \node[] (z) at (1.5, -1)   {$Z$};
    \node[lv] (u) at (4.5, 0)   {$U$};
    \path[->] (a) edge node [above right] {} (m);
    \path[->] (m) edge node [above right] {} (y);
    \path[->] (z) edge node [above right] {} (y);
    \path[->] (a) edge node [above left] {} (z);
    \path[->,dashed] (u) edge node [above right] {} (m);
    \path[->,dashed] (u) edge node [above right] {} (y);
  \end{tikzpicture}
  \\
  \vspace{0.3cm}
  \begin{tikzpicture}[>=stealth,scale=0.7,thick]
    \node[] (title) at (-0.3, 1)   {(C)};
    \node[] (a) at (0, 0)   {$A$};
    \node[] (m) at (3, 0)   {$M$};
    \node[] (y) at (3, -2)   {$Y$};
    \node[] (z) at (1.5, -1)   {$Z$};
  	\node[] (l) at (1.5, 0)   {$L$};
    \node[lv] (u) at (4.5, 0)   {$U$};
    \path[->] (a) edge node [above right] {} (l);
    \path[->] (l) edge node [above right] {} (m);
    \path[->] (m) edge node [above right] {} (y);
    \path[->] (z) edge node [above right] {} (y);
    \path[->] (a) edge node [above left] {} (z);
    \path[->,dashed] (u) edge node [above right] {} (m);
    \path[->,dashed] (u) edge node [above right] {} (y);
  \end{tikzpicture}
  \hspace{0.3cm}
  \begin{tikzpicture}[>=stealth,scale=0.7,thick]
    \node[] (title) at (-0.3, 1)   {(D)};
    \node[] (a) at (0, 0)   {$A$};
    \node[] (m) at (3, 0)   {$M$};
    \node[] (y) at (3, -2)   {$Y$};
    \node[] (z) at (1.5, -1)   {$Z$};
    \node[] (l) at (1.5, 0)   {$L$};
    \node[lv] (u) at (4.5, 0)   {$U$};
    \path[->,line width=2pt] (a) edge node [above right] {} (l);
    \path[->] (l) edge node [above right] {} (m);
    \path[->] (m) edge node [above right] {} (y);
    \path[->] (z) edge node [above right] {} (y);
    \path[->,line width=2pt] (a) edge node [above left] {} (z);
    \path[->,dashed] (u) edge node [above right] {} (m);
    \path[->,dashed] (u) edge node [above right] {} (y);
  \end{tikzpicture}
	\caption{Hidden variable causal DAGs with with mediating instruments $L$ for the path $A\rightarrow M$ (A),
	$Z$ for the path $A\rightarrow Y$ (B), and a combination of both (C).
  The DAG in panel (D) is an extended deterministic graph of the DAG in Figure~\ref{exampleDAG1}A, with thick edges indicating a deterministic relationship. \label{exampleDAG3}}
\end{figure}

Cross-world independence~\eqref{ccwass} is violated in the causal DAG $\mathcal{G}(\mathbf{V}\cup\mathbf{H})$ with $\mathbf{V} = \{A,L,M,Y\}$ and $\mathbf{H} = \{U\}$, in Figure~\ref{exampleDAG3}A, because of unmeasured mediator-outcome confounding.
Nonetheless, by exploiting both conditional independencies that are naturally encoded in $\mathcal{G}(\mathbf{V}\cup\mathbf{H})$
and conditional counterfactual independencies implied by the following NPSEM representation of $\mathcal{G}(\mathbf{V}\cup\mathbf{H})$
\begin{align*}
    L(a') &:= f_L(a',\epsilon_L)\\
    M(l) &:= f_M(l,U,\epsilon_M)\\
    Y(a,m) &:= f_Y(a,m,U,\epsilon_Y),
\end{align*}
$p(Y(a,M(a'))=y)$ can be identified from observed data as follows
\begin{align}
p(Y(a,M(a'))=y) &= p(Y(a,M(L(a')))=y)\nonumber\\
&= \sum_{l,m} p(Y(a,m)=y, M(l)=m, L(a')=l)\nonumber\\
&= \sum_{u,l,m} p(Y(a,m)=y\vert u) 
p(M(l)=m\vert u)p(L(a')=l)p(u) \nonumber\\
&= \sum_{u,l,m} p(y\vert a,m,u) p(m\vert l, u) p(l\vert a') p(u)\nonumber\\
&= \sum_{u,l,m} p(y\vert a,l,m,u) p(m\vert a,l,u) p(l\vert a') p(u\vert a, l)\nonumber\\
&= \sum_{l} p(y\vert a,l) p(l\vert a')\label{medinstrres}.
\end{align}
This result should not come as a big surprise: since the effect of treatment on the mediator is (assumed to be) entirely mediated by $L$,
and, in addition, $L$ only affects the outcome via $M$, $L$ can simply substitute for $M$.
However, the sufficient conditions outlined so far (especially assumption~\eqref{ccwass}) do not naturally lead to this simple result.

In the remainder of this section, we therefore take a step back.
Armed with the tools and concepts from the previous chapter, we take a closer look at the commonalities
that characterize the key problems in the examples in sections~\ref{unmeasuredconf} and~\ref{intermediateconf}.
The resulting insights offer a framework that allows to extend complete algorithms for identification of total causal effects
(see sections 1.3.4 to 1.3.6 of the previous chapter) to mediation settings.
Moreover, this extension has produced a complete graphical criterion for identification under NPSEMs,
not only of natural direct and indirect effects, but of path-specific effects in general \citep{Shpitser2013}.
In sections~\ref{implcompl} and~\ref{conccl}, we highlight that this graphical criterion gives rise to complementary identification strategies
that may, in addition, help to shed new light on an ongoing debate about the controversial cross-world nature of path-specific effects.

\subsection{Building blocks for complete graphical identification criteria \label{buildblocks}}
Key to violation of assumptions~\eqref{pc1} and~\eqref{pc2} in the examples in sections~\ref{unmeasuredconf} and~\ref{intermediateconf}, respectively,
is the occurrence of conflicting treatment assignments in certain factors of the identifying functional. 
It can be shown that this problem arises whenever the conflict is situated in factors involving distributions of ancestors of the outcome --- which, by convention, include the outcome itself ---
that belong to a common \textit{confounded component} (abbreviated: \textit{c-component}) \citep{Tian2003} or \textit{district} \citep{Richardson2003}. 
In this subsection, we first provide the necessary conceptual background on districts, district factorization and complete identification algorithms for total treatment effects, 
as discussed in more technical detail in the previous chapter.

\paragraph{Confounded paths, districts and district factorizations}
Following \citep{Shpitser2008}, we first define a \emph{confounded path} to be a path where all directed arrowheads point at observed nodes, and never away from observed nodes.
To avoid cluttering causal DAGs $\mathcal{G}(\mathbf{V}\cup\mathbf{H})$ with large numbers of hidden variables, unobserved common causes of any two observed nodes are often omitted, while their presence is, instead, indicated by
bidirected edges ($\leftrightarrow$).
This latent projection operation, as discussed in previous chapters, gives rise to \emph{acyclic directed mixed graphs} (ADMGs) \citep{Richardson2003}, which contain only observed nodes $\mathbf{V}$ and both directed and bidirected edges (see section 1.3 of chapter 2).
These graphs encode conditional independencies between observed variables via m-separation \citep{Richardson2003}, a graphical criterion closely related to d-separation for causal DAGs containing only directed edges.
Throughout, we will denote latent projection ADMGs of hidden variable causal DAGs $\mathcal{G}(\mathbf{V}\cup\mathbf{H})$ by $\mathcal{G}(\mathbf{V})$.
In ADMGs, confounded paths can similarly be defined as paths that contain only bidirected edges.

A district may now be defined as the maximal set of observed nodes that are pairwise connected by confounded paths.
The set of all observed nodes can thus be partitioned into disjoint districts $\mathbf{S} \in \mathcal{D}(\mathcal{G}(\mathbf{V}))$,
where $\mathcal{D}(\mathcal{G}(\mathbf{V}))$ denotes the set of districts in the latent projection $\mathcal{G}(\mathbf{V})$.
Each of these districts consists of either a single observed node or a set of observed nodes that are pairwise connected by confounded paths.
The importance of districts may be appreciated by the fact that their disjointness implies that the marginal distribution
of observed variables $p(\mathbf{x_V})$ in $\mathcal{G}(\mathbf{V}\cup \mathbf{H})$ factorizes as the product of their corresponding \emph{kernels} or \emph{c-factors}
$$p(\mathbf{x_V}) = \prod_{\mathbf{S} \in \mathcal{D}(\mathcal{G}(\mathbf{V}))} Q[\mathbf{S}],$$
where each kernel corresponds to
\begin{align*}
  Q[\mathbf{S}] = \sum_{\mathbf{x}_{\text{u}_\mathcal{G}(\mathbf{S})}} \prod_{V \in \mathbf{S}} p(x_V\vert \mathbf{x}_{\text{pa}_{\mathcal{G}}(V)}, \mathbf{x}_{\text{u}_{\mathcal{G}}(V)}) p(\mathbf{x}_{\text{u}_{\mathcal{G}}(\mathbf{S})}),
\end{align*}
and where $\text{pa}_\mathcal{G}(V)$ and $\text{u}_\mathcal{G}(V)$ denote the set of observed and unobserved parents of $V$ in $\mathcal{G}(\mathbf{V}\cup \mathbf{H})$, respectively.
For each $Q[\mathbf{S}]$, the product runs across all observed nodes $V \in \mathbf{S}$
and the summation over all possible realisations of unobserved parents of $V \in \mathbf{S}$.

For instance, in the causal DAG $\mathcal{G}(\mathbf{V}\cup\mathbf{H})$ with $\mathbf{V} = \{A,M,Y\}$ and $\mathbf{H} = \{U\}$, in Figure~\ref{exampleDAG1}A,
$M$ and $Y$ are connected by a confounded path as they share an unmeasured parent $U$.
The set of observed variables can thus be partitioned into two districts: $\{A\}$ and $\{M,Y\}$.
Moreover, $p(a,m,y)$ factorizes as the product of the kernels of these districts:
$$p(a,m,y) = Q[\{A\}]Q[\{M,Y\}] = p(a) \sum_u p(y\vert a,m,u) p(m\vert a,u) p(u).$$

Tian and Pearl \citep{Tian2003} pointed out that every $Q[\mathbf{S}]$ can be interpreted as $p(\mathbf{x_S}\vert \text{do}(\mathbf{x}_{\text{pa}_{\mathcal{G}}(\mathbf{S})}))$,
the distribution of $\mathbf{S}$ under an intervention that sets all its observed parents $\text{pa}_\mathcal{G}(\mathbf{S})$ to $\mathbf{x}_{\text{pa}_\mathcal{G}(\mathbf{S})}$.
Moreover, they proved that every $Q[\mathbf{S}]$, for which $\mathbf{S} \in \mathcal{D}(\mathcal{G}(\mathbf{V}))$, is identifiable from observed data.
For example, since $A\ci U$ in Figure~\ref{exampleDAG1}A, it is relatively trivial to show that $$Q[\{M,Y\}] = \sum_u p(y\vert a,m,u) p(m\vert a,u) p(u) = p(y\vert a,m) p(m\vert a).$$
We next illustrate that districts (and their corresponding kernels) form the building blocks of complete graphical identification algorithms for total causal effects \citep{Huang2006,Shpitser2006a,Tian2003},
since interventional distributions can always be expressed as the marginal of a truncated district factorization.

\paragraph{Truncated district factorizations in hidden variable causal DAGs}
In hidden variable causal DAGs $\mathcal{G}(\mathbf{V}\cup\mathbf{H})$, observed nodes $V\in\mathbf{V}$ that are connected by confounded paths group together into districts $\mathbf{S} \in \mathcal{D}(\mathcal{G(\mathbf{V})})$.
Consequently, factorizations of $p(\mathbf{x_V})$ involve kernels that can be interpreted as multivariate interventional distributions.
Because district factorizations extend usual Markov factorizations for causally sufficient DAGs to hidden variable causal DAGs with respect to $p(\mathbf{x_V})$,
this permits us to express $p(Y(a)=y)$ as the marginal of a truncated version of the district factorization of $p(\mathbf{x_V})$:
\begin{align}\label{gform2}
    p(Y(a)=y) &= \sum_{\mathbf{x}_{\mathbf{Y^*}\setminus Y}} \prod_{\mathbf{D} \in \mathcal{D}(\mathcal{G}(\mathbf{V})_\mathbf{Y^*})} p(\mathbf{x_D} \vert \text{do}(a_{\text{pa}_{\mathcal{G}}(\mathbf{D})\cap A}, \mathbf{x}_{\text{pa}_{\mathcal{G}}(\mathbf{D})\setminus A})),
\end{align}
where $a_{\text{pa}_{\mathcal{G}}(\mathbf{D})\cap A} = a$ if there exists a directed path of the form $A\rightarrow D\rightarrow ...\rightarrow Y$ (with $D\in\mathbf{D}$), $a_{\text{pa}_{\mathcal{G}}(\mathbf{D})\cap A} = \emptyset$ if no such path exists,
and where $\mathbf{Y^*} \equiv \text{an}_{\mathcal{G}_{\mathbf{V}\setminus A}}(Y)$ denotes the set of ancestors of $Y$ (including $Y$) in a subgraph $\mathcal{G}(\mathbf{V})_{\mathbf{V}\setminus A}$ of the latent projection $\mathcal{G}(\mathbf{V})$.
Here, the product runs across all districts $\mathbf{D} \in \mathcal{D}(\mathcal{G(\mathbf{V})}_{\mathbf{Y^*}})$ in a subgraph of $\mathcal{G(\mathbf{V})}$,
and the summation is made over all possible realisations of the nodes in these districts, except for the outcome.

Equation~\ref{gform2} indicates that the original problem of identifiability of $p(Y(a)=y)$ can be reduced to a set of smaller identification problems within a subgraph of $\mathcal{G}(\mathbf{V}\cup\mathbf{H})$.
Logically, each district $\mathbf{D} \in \mathcal{D}(\mathcal{G(\mathbf{V})}_{\mathbf{Y^*}})$ is a subset of a district $\mathbf{S} \in \mathcal{D}(\mathcal{G(\mathbf{V})})$.
Even though each kernel $Q[\mathbf{S}]$ is identifiable from observed data, identification of some kernels $Q[\mathbf{D}]$ may fail under certain hidden variable causal models.
In fact, $p(Y(a)=y)$ is identifiable if and only if every kernel $Q[{\mathbf{D}}]$ in Equation~\ref{gform2} can be recovered from observed data.
In the absence of hidden variables, $\mathcal{D}(\mathcal{G}(\mathbf{V})_{\mathbf{Y^*}}) = \mathcal{D}(\mathcal{G}(\mathbf{V}))$,
such that $p(Y(a)=y)$ is always identifiable from observed data and Equation~\ref{gform2} reduces to the well-known g-formula \citep{Robins1986}.

For instance, in $\mathcal{G}(\mathbf{V}\cup\mathbf{H})$ in Figure~\ref{exampleDAG1}A,
$\mathbf{Y^*} = \mathbf{V}\setminus A = \{M,Y\}$, such that there is only one district $\{M,Y\}$ in $\mathcal{G(\mathbf{V})}_{\mathbf{Y^*}}$.
Its corresponding kernel $Q[\{M,Y\}]$ in Equation~\ref{gform2} perfectly corresponds
to a kernel in the latent projection $\mathcal{G}(\mathbf{V})$, such that it can readily be identified from observed data.
Consequently, $p(Y(a)=y)$ is identified by
\begin{align}\label{MYconfid}
  \sum_{m} p(y,m\vert \text{do}(a)) = \sum_{u,m} p(y\vert a,m,u) p(m\vert a,u) p(u) = \sum_{m} p(y\vert a,m) p(m\vert a) = p(y\vert a).
\end{align}

In contrast to this simple example, truncation of district factorizations can become much more complicated under hidden variable DAGs.
That is, whenever a district $\mathbf{D} \in \mathcal{D}(\mathcal{G(\mathbf{V})}_{\mathbf{Y^*}})$ is a \emph{proper} subset of a district $\mathbf{S} \in \mathcal{D}(\mathcal{G}(\mathbf{V}))$,
identification of $Q[\mathbf{D}]$ requires (repeated) application of the `fixing operation' (as described in the previous chapter) and may eventually fail.\footnote{Application of the `fixing operation' in a conditional ADMGs (as described in the previous chapter)
is essentially equivalent to the systematic removal of certain non-essential nodes in subgraphs of the latent projection ADMG $\mathcal{G}(\mathbf{V})$.}
Problematic graphical structures that hinder identification under hidden variable causal models have been discussed in detail in \citep{Shpitser2006}.

\subsection{The central notion of recantation}
Having provided the necessary theoretical and conceptual background on complete identification algorithms for total treatment effects,
we now pick up where we left off in the beginning of section~\ref{buildblocks} and provide more formality and generality by introducing the central notion of \emph{recantation}.
This notion will enable us to map crucial cross-world independencies onto an intuitive graphical criterion under NPSEMs.
It can therefore be viewed to serve as a passkey that permits easy translations from cross-world quantities, used to define path-specific effects, to `single-world' interventional quantities.
Identification of the latter can then simply be passed on to well-established algorithms for identifying total causal effects.
Ultimately, when combined with such complete algorithms, this graphical criterion does not only delineate complete identification criteria for natural effects,
but also for more generally defined path-specific effects.

\subsubsection{The recanting witness criterion \label{recantwitnesssection}}
Cross-world counterfactual $Y(a,M(a'))$ in Figure~\ref{exampleDAG2}A corresponds to $Y(a,L(a),M(a',L(a')))$
and thus represents the response of the outcome to two hypothetical interventions which
set to $A$ to $a$ for the purpose of $\pi = \{A\rightarrow Y; A\rightarrow L\rightarrow Y\}$, on the one hand,
and to $a'$ for the purpose of $\overline\pi = \{A\rightarrow M\rightarrow Y; A\rightarrow L\rightarrow M\rightarrow Y\}$, on the other hand.
However, even though the interventional distribution $p(Y(a)=y)$ is identified by
\begin{align*}
  \sum_{l,m} p(y\vert \text{do}(a,l,m)) p(m\vert \text{do}(a,l)) p(l\vert \text{do}(a)) = \sum_{l,m} p(y\vert a,l,m) p(m\vert a,l) p(l\vert a) = p(y\vert a),
\end{align*}
identification of $p(Y(a,M(a'))=y)$ is hampered because of conflicting treatment assignments $a$ and $a'$ in the single node district $L$.
Here, $L$ is called a \textit{recanting witness} \citep{Avin2005}, for the following reason.
Identification of the natural indirect effect via $M$ requires a first statement from $L$ that
\textit{blocks} the path $A\rightarrow L\rightarrow Y$ in order to keep treatment from transmitting its effect on the outcome other than through $M$ (as this path is regarded part of the natural direct effect).
However, $L$ subsequently needs to retract this statement in favor of a new statement
which \textit{refrains from blocking} the path $A\rightarrow L\rightarrow M\rightarrow Y$ in order to allow treatment to transmit its entire effect on the mediator (as blocking would imply adjusting away part of the natural indirect effect).
Clearly, we can't have it both ways.

The \emph{recanting witness criterion} \citep{Avin2005} formalizes this requirement of having no such witnesses along $\pi$ to enable identification of the $\pi$-specific effect.
More specifically, a child $L$ of treatment $A$ is called a recanting witness\footnote{The definition we provide here is restricted to $\pi$-specific effects from $A$ to $Y$, with both $A$ and $Y$ being singletons.
A more general definition requires making reference to \emph{proper causal paths}, and is given in the previous chapter.}
for $\pi$ (and by symmetry, also for $\overline\pi$) if there exists a directed path in $\pi$ of the form $A\rightarrow L\rightarrow ...\rightarrow Y$
and another directed path in $\overline\pi$ of the form $A\rightarrow L\rightarrow ...\rightarrow Y$.
Avin, Shpitser and Pearl \citep{Avin2005} demonstrated that if and only if there is no recanting witness for $\pi$ in a causal DAG $\mathcal{G}(\mathbf{V})$ without hidden variables,
then the $\pi$-specific effect is identified under the NPSEM representation of $\mathcal{G}(\mathbf{V})$.
More specifically, the distribution of the corresponding nested counterfactual $Y(\pi,a,a')$ is then identified from observed data as
\begin{align}\label{edgegform}
    p(Y(\pi, a, a')=y) &= \sum_{\mathbf{x}_{\mathbf{V}\setminus (A\cup Y)}} \prod_{V \in \mathbf{V}\setminus A} p(x_V\vert a_{\text{pa}^{\pi}_{\mathcal{G}}(V)\cap A}, a^{\prime}_{\text{pa}^{\overline\pi}_{\mathcal{G}}(V)\cap A}, \mathbf{x}_{\text{pa}_{\mathcal{G}}(V)\setminus A}),
\end{align}
where $\text{pa}^{\pi}_{\mathcal{G}}(V)$ denotes the set of parents of $V$ in $\mathcal{G}(\mathbf{V})$ along an edge which is part of a path in $\pi$.
This result has been referred to as the \emph{edge g-formula} \citep{Shpitser2016} because it generalizes the ordinary g-formula \citep{Robins1986}
in that it permits different treatment assignments along separate sets of edges $A\rightarrow V$.
More specifically, in different Markov factors, treatment $A$ is set to either $a$ or $a'$ depending on whether or not $\{A\rightarrow V\rightarrow ...\rightarrow Y\} \in \pi$.
The recanting witness criterion thus implicitly imposes the restriction that a single edge $A\rightarrow V$ can only be assigned a single treatment value or,
in other words, that treatment assignment must be \emph{edge consistent} for $p(Y(\pi, a, a')=y)$ to be identified \citep{Shpitser2016}.
The mediation formula (Equation~\ref{medform}) can be viewed as a specific case of this more general identifying functional.

\subsubsection{The recanting district criterion \label{recantdistrictsection}}
Even though the recanting witness criterion gives a complete characterization of settings when path-specific effects are identified under DAGs without hidden variables,
it does not suffice as a graphical identification criterion in hidden variable DAGs.
This can be seen from the illustrating example in section~\ref{exampleDAG1}, which suffers from unmeasured mediator-outcome confounding.
For instance, even though $p(Y(a)=y)$ is identified by Equation~\ref{MYconfid} under the hidden variable causal DAG $\mathcal{G}(\mathbf{V}\cup\mathbf{H})$ in Figure~\ref{exampleDAG1}A,
$p(Y(a,M(a'))=y)$ is not identified by $\sum_{m} p(y\vert a,m) p(m\vert a')$, simply because we cannot readily integrate out $U$ under conflicting treatment assignments $a$ and $a'$.
Identification of $p(Y(a,M(a'))=y)$ is thus hindered because of conflicting treatment assignments $a$ and $a'$ in the district $\{M,Y\}$,
where $A$ is set to $a$ for the purpose of $\pi = \{A\rightarrow Y\}$
and to $a'$ for the purpose of $\overline\pi = \{A\rightarrow M\rightarrow Y\}$.

Inspired by complete algorithms for identifying $p(Y(a)=y)$, based on district factorizations,
Shpitser \citep{Shpitser2013} recently extended the recanting witness criterion to hidden variable DAG settings.
This extension is conceptually fairly simple. As districts rather than single nodes are the building blocks of the factorization of $p(\mathbf{x_V})$,
the term `witness' simply needs to be replaced by the term `district'.
Informally, this extended criterion requires there to be no `conflict of interest' between members of a common district within a particular subgraph $\mathcal{G}(\mathbf{V})_{\mathbf{Y^*}}$.
Formally, a district $\mathbf{D} \in \mathcal{D}(\mathcal{G}(\mathbf{V})_{\mathbf{Y^*}})$ is said to be a \emph{recanting district} for $\pi$
if there exists a directed path in $\pi$ of the form $A\rightarrow D\rightarrow ...\rightarrow Y$ as well as a directed path in $\overline\pi$ of the form $A\rightarrow D'\rightarrow ...\rightarrow Y$,
where $D, D' \in \mathbf{D}$ (and possibly $D = D'$).

Shpitser \citep{Shpitser2013}, moreover, showed that only in the absence of a recanting district for $\pi$,
the cross-world counterfactual distribution $p(Y(\pi,a,a')=y)$ can be expressed as a functional of interventional distributions
\begin{align}\label{edgegform2}
    p(Y(\pi, a, a')=y) &= \sum_{\mathbf{x}_{\mathbf{Y^*}\setminus Y}} \prod_{\mathbf{D} \in \mathcal{D}(\mathcal{G}_{\mathbf{Y^*}})} p(\mathbf{x_D} \vert \text{do}(a_{\text{pa}^{\pi}_{\mathcal{G}}(\mathbf{D})\cap A}, a^{\prime}_{\text{pa}^{\overline\pi}_{\mathcal{G}}(\mathbf{D})\cap A}, \mathbf{x}_{\text{pa}_{\mathcal{G}}(\mathbf{D})\setminus A})).
\end{align}
This functional is a generalization of the truncated district factorization used to identify $p(Y(a)=y)$ in hidden variable causal DAGs (Equation~\ref{gform2}) via Tian's \textbf{ID} algorithm \citep{Tian2003}.
Equation~\ref{edgegform2} closely matches the formulation in Equation~\ref{edgegform} in that it also permits different treatment assignments along separate sets of edges,
while replacing single nodes $V$ by districts $\mathbf{D} \in \mathcal{D}(\mathcal{G}(\mathbf{V})_{\mathbf{Y^*}})$.
It can therefore also be considered an extension of the edge g-formula to hidden variable causal DAGs.
However, an additional condition must hold to enable further translation of the involved interventional quantities
onto observable quantities. That is, whereas each of the kernels in Equation~\ref{edgegform} is identifiable from observed data in causal DAGs without hidden variables,
identifiability of kernels in Equation~\ref{edgegform2} is not guaranteed because of the assumed presence of hidden variables.
Nonetheless, if $p(Y(a)=y)$ is identified via Tian's \textbf{ID} algorithm,
this logically implies that each kernel in the above functional is expressible in terms of observed data.
Importantly, the recanting district criterion thus needs to be complemented by identifiability of the total causal effect $p(Y(a)=y)$
to give a complete characterization of identification conditions for path-specific effects under hidden variable causal DAGs (interpreted as NPSEMs) \citep{Shpitser2013}.
Whereas the recanting district criterion enables translations from cross-world counterfactual quantities into `single world' interventional quantities,
the \textbf{ID} algorithm then verifies whether these interventional quantities can be expressed as functionals of the observed data.

\paragraph{A somewhat more involved example revisited}
Consider again $\mathcal{G}(\mathbf{V}\cup\mathbf{H})$ in Figure~\ref{Pearl2014fig5fsingle}A, along with the subgraphs of interest $\mathcal{G}(\mathbf{V})_{\mathbf{V}\setminus A}$ and $\mathcal{G}(\mathbf{V})_{\mathbf{Y^*}}$.
Since there is no recanting district for the set of pathways that capture the natural (in)direct effect in $\mathcal{G}(\mathbf{V})_{\mathbf{Y^*}}$,
the identifying functional for $p(Y(a,M(a'))=y)$ can be expressed as
\begin{align} \label{invexintv}
  &\sum_{c_1,c_3,m} p(y\vert do(a,m,c_3)) p(m\vert do(a',c_1)) p(c_3\vert do(c_1)) p(c_1)
\end{align}
by application of Equation~\ref{edgegform2}. Moreover, because $p(Y(a)=y)$ is identified by
\begin{align} \label{invextotal}
  &\sum_{c_1,c_2,c_3,m} p(y\vert a,m,c_3) p(m\vert a,c_1,c_2) p(c_1) p(c_2) p(c_3\vert c_1)
\end{align}
via Tian's \textbf{ID} algorithm, we can re-express Equation~\ref{invexintv} as Equation~\ref{invexnatural}
by simply plugging in appropriate treatment assignments in the respective factors of Equation~\ref{invextotal}.

Note that Pearl's `divide and conquer' approach, as discussed in section~\ref{idbeyondadjcrit}, required searching the space of candidate covariate sets $\mathbf{C}$
that not only satisfy cross-world indendence~\eqref{ccwass} but also conditions~\eqref{mac} and~\eqref{yamc}.
Shpitser's identification approach is therefore not only (more) complete,
but arguably also more insightful as it clarifies that the main difficulty in identifying $p(Y(a,M(a'))=y)$ is identification of the total treatment effect,
which involves repeated application of the fixing operator.

\subsubsection{A new perspective on cross-world independence}
From the perspective of the recanting district criterion, the need for an observed covariate set $\mathbf{C}$ that is sufficient to adjust for confounding of the mediator-outcome relation
(given treatment) serves to establish that mediator and outcome belong to separate districts so that conflicting treatment assignments causes no further identification problems
(provided that no member of $\mathbf{C}$ is affected by treatment).
For instance, in Figures~\ref{exampleDAG1}B and C, a sufficient adjustment set $\{C\}$ enables to pull apart the district $\{M,Y\}$ and resolve the conflict in order to ensure
the validity of cross-world assumption~\eqref{ccwass} that permits factorizing $p(Y(a,m)=y, M(a')=m\vert \mathbf{c})$ as $p(Y(a,m)=y\vert \mathbf{c}) p(M(a')=m\vert \mathbf{c})$.

Importantly, the central notion of recantation thus groups Pearl's graphical criteria~\eqref{pc1} and~\eqref{pc2}
for establishing cross-world independence~\eqref{ccwass} under NPSEMs
by offering a framework that allows their respective violations to be interpreted as distinct instances of essentially the same problem.
As will be discussed in the next section, 
the implications of this graphical criterion reach beyond Pearl's \citep{Pearl2001} sufficient conditions, as discussed in section~\ref{id1}.

\section{Complementary identification strategies \label{implcompl}}
The completeness of Shpitser's \cite{Shpitser2013} new identification approach reveals that Pearl's \cite{Pearl2001} conditions~\eqref{ccwass}-\eqref{mac}-\eqref{yamc} may not be necessary to recover natural effects from observed data.
That is, simply combining cross-world independence~\eqref{ccwass} and identifiability of the total causal effect may suffice to identify $p(Y(a,M(a'))=y)$.
When cross-world independence~\eqref{ccwass} can thus be established upon adjustment for a covariate set $\mathbf{C}$, one may simply assess identifiability of $p(Y(a)=y)$ via the \textbf{ID} algorithm
instead of assessing identifiability of both $p(M(a)=m\vert \mathbf{c})$ and $p(Y(a,m)=y\vert \mathbf{c})$ via the more complicated \textbf{IDC} algorithm.

Interestingly, the completeness of this novel result also highlights that, in some rare cases,
cross-world independence~\eqref{ccwass} --- despite being a sufficient condition for `experimental' identification \citep{Pearl2001} --- may not be required either.
This was already exemplified by the case of Figure~\ref{exampleDAG3}A, which we now revisit.

\subsection{Interchanging cross-world assumptions}
Careful inspection of the hidden variable causal DAG $\mathcal{G}(\mathbf{V}\cup\mathbf{H})$ in Figure~\ref{exampleDAG3}A yields that $\mathbf{Y^*} = \{L,M,Y\}$,
such that the subgraph $\mathcal{G}(\mathbf{V})_{\mathbf{Y^*}}$ can be partitioned into districts $\{L\}$ and $\{M,Y\}$.
The absence of a recanting district for $\pi = \{A\rightarrow Y\}$, which transmits the natural direct effect with respect to $M$,
and identifiability of $p(Y(a)=y)$, by randomization of treatment, then leads to the same identification result for $p(Y(a,M(a'))=y)$ as obtained in Equation~\ref{medinstrres},
but derived more elegantly via application of Equation~\ref{edgegform2}:
\begin{align*}
	\sum_{l,m} p(y,m\vert \text{do}(a,l)) p(l\vert \text{do}(a')) &= \sum_{l,m} p(y\vert a,l,m) p(m\vert a,l) p(l\vert a') = \sum_{l} p(y\vert a,l) p(l\vert a').
\end{align*}
This result can be explained by the fact that, in this case, the recanting district criterion does not serve to establish identifiability
via cross-world independence~\eqref{ccwass},
but, instead, via the alternative cross-world independence assumption $Y(a,m) \ci L(a')$ encoded in the NPSEM representation of $\mathcal{G}(\mathbf{V}\cup\mathbf{H})$.
Indeed, the derivations in Equation~\ref{medinstrres} illustrate that the \textit{mediating instrument} $L$ achieves to prevent the conflict between treatment assignments $a$ and $a'$
from taking place \emph{within} the district $\{M,Y\}$ by diverting treatment state $a'$ to itself, thereby fulfilling its mediating role, literally and figuratively.
A crucial insight here is that when $L$ is assumed to mediate the entire treatment effect on the mediator $M$, then the latter is no longer a child of $A$,
and hence, cannot receive any input from $A$ that may conflict with input to other children of $A$ in the same district.

A mediating instrument on the path between treatment and outcome, such as $Z$ in Figure~\ref{exampleDAG3}B, would similarly allow to make progress
upon substituting~\eqref{ccwass} by cross-world independence $Z(a) \ci M(a')$.

\subsection{Two types of auxiliary variables \label{auxvar}}
The above examples illustrate that, when $p(Y(a)=y)$ is identifiable, further identification of $p(Y(a,M(a'))=y)$ by the recanting district criterion
can be achieved under NPSEMs with the aid of two types of auxiliary variables.
Each type can be viewed to have its own distinct strategy for preventing recantation.\footnote{Note that this classification is analogous to the one often used for auxiliary variables that aid identification of treatment effects,
where identification can be achieved via two main strategies: using either the back-door criterion (i.e. standard adjustment for covariates) or the front-door criterion (i.e. sequential adjustment by means of a mediating instrument).}

The first type, such as $C$ in Figures~\ref{exampleDAG1}B and~\ref{exampleDAG1}C, aims to prevent conflicting treatment assignments within districts by separating nodes of a common district, such as $\{M,Y\}$ in Figure~\ref{exampleDAG1}A,
which is bound to recant due to unmeasured mediator-outcome confounding, into different districts.
Adjustment for this type of covariates specifically aims to strengthen cross-world assumption~\eqref{ccwass}.
The second type, such as $L$ or $Z$ in Figures~\ref{exampleDAG3}A,~\ref{exampleDAG3}B and~\ref{exampleDAG3}C, avoids conflicts in a specific district (such as $\{M,Y\}$) not by separating its nodes into different districts,
but instead hosting one potential `troublemaker' in its own district.
Such mediating instruments therefore do not aspire to establish assumption~\eqref{ccwass},
but instead target identification by means of alternative cross-world assumptions that may substitute for assumption~\eqref{ccwass}.

This result is important because mediating instruments, while useful to identify $p(M(a)=m\vert \mathbf{c})$ and/or $p(Y(a,m)=y\vert \mathbf{c})$ in order to satisfy conditions~\eqref{mac} and~\eqref{yamc},
cannot aid in (avoiding recantation by) establishing cross-world independence~\eqref{ccwass} because this can only be achieved by means of covariate adjustment \citep{Pearl2014}.
The recanting district criterion reveals the extended utility of mediating instruments as auxiliary variables that may,
nonetheless, help to avoid recantation by establishing cross-world independencies that substitute for cross-world independence~\eqref{ccwass}.

\subsection{Mediating instruments --- some reasons for skepticism}
Contrary to the long-held belief that identification of $p(Y(a,M(a'))=y)$ hinges on the assumption that no mediator-outcome confounding is left unadjusted,
mediating instruments arm us with additional identification power in the presence of such unmeasured confounding.
This provides researchers different identification strategies, each relative to a specific set of assumptions.
One may use this as a basis for a sensitivity analysis, or adopt the strategy that corresponds with the most plausible assumptions given a certain research context.
However, some caution is warranted.

First, the recanting district criterion indicates that, when resorting to mediating instruments, the requirement of no unmeasured confounding is simply shifted
from the mediator-outcome relation to both the instrument-mediator and instrument-outcome relations.
This can be seen upon noting that either type of unmeasured confounding results in the instrument
being `absorbed' into the district $\{M,Y\}$, such that it expands to $\{L,M,Y\}$ in Figure~\ref{exampleDAG3}A or to $\{Z,M,Y\}$ in Figure~\ref{exampleDAG3}B, respectively.
Both of these would, however, be recanting with respect to the set of pathways $\pi$ that transmit either the natural direct or natural indirect effect.
As for mediator-outcome confounding, neither types of unmeasured confounding can be avoided by treatment randomization.

Second, the assumption that $L$ or $Z$ is a mediating instrument involves strong and often unrealistic exclusion restrictions.\footnote{Importantly, $L$ or $Z$ may also correspond to covariate sets (rather than being singletons) that satisfy the stated conditions.}
For instance, for $L$ in Figure~\ref{exampleDAG3}A to be a mediating instrument, it would need to mediate the entire effect of $A$ on $M$.
Despite being a strong assumption, it is partially testable from observed data when there is no unmeasured $L-M$ or $A-M$ confounding, as in Figure~\ref{exampleDAG3}A, for then $M$ must be conditionally independent of $A$, given $L$.
Likewise, for $Z$ in Figure~\ref{exampleDAG3}B to be a mediating instrument for the path $A\rightarrow Y$,
it would need to mediate the entire direct treatment effect on the outcome that is not mediated by $M$.
However, the requirement that $Z$ and $M$ together mediate the entire treatment effect, which implies $Y\ci A\vert \{Z,M\}$, is untestable in the presence of unmeasured mediator-outcome confounding.
Even though these assumptions cannot directly be verified from observed data,
in principle, along with the aforementioned unmeasured confounding assumptions, they lend themselves to experimental verification.

Third, mediating instruments do not resolve the previously considered identification problems in the presence of treatment-induced mediator-outcome confounding by a recanting witness.
For instance, the exclusion restriction that $L$ does not directly affect $Y$ in Figure~\ref{exampleDAG3}A, or that
$Z$ does not affect $M$ in Figure~\ref{exampleDAG3}B, can be thought of as a constraint that prevents the instrument from turning into a recanting witness.

\section{From mediating instruments to conceptual clarity \label{conccl}}
Even though the practical use of mediating instruments, as an alternative route to identification of natural effects, may be debatable, their added value is more immediate on a conceptual level.
Such instruments may help to frame some recent conceptual development that aims to cast mediation analysis into a more strict interventionist paradigm, void of untestable cross-world assumptions \citep{Robins2010}.
Before going on to discuss this development, we briefly sketch some difficulties that may arise when interpreting natural effects, at least from an interventional point of view.

\subsection{In search of operational definitions}
When it comes to the interpretation of natural direct effects, critics adhering to the slogan `no causation without manipulation' have repeatedly emphasized
the operational question of \emph{how} exactly one may go about blocking the treatment's effect on the mediator, in order to recover $M(0)$ in treated subjects, without affecting the direct path from treatment to outcome \citep[e.g.][]{Didelez2006}.
Inevitably, any answer to this question invokes a mediating instrument, such as $L$ in Figure~\ref{exampleDAG3}A, that can be intervened on in order to prevent treatment from exerting its effect on the mediator.
Likewise, it is difficult to imagine an intervention that would block only the direct path from treatment to outcome, without conceptualizing a mediating instrument such as $Z$ in Figure~\ref{exampleDAG3}B.

\subsection{Deterministic expanded graphs \label{detexpgraphs}}
It thus seems that mediating instruments provide some sort of necessary extension to the original causal diagram that permits interventionist interpretations of natural effects.
The conceptual notion of an expanded graph with two mediating instruments, as depicted in Figure~\ref{exampleDAG3}C,
corresponds very closely to what has been described by Robins and Richardson \cite{Robins2010}.
In settings where $L$ and $Z$ can confidently be considered mediating instruments,
the recanting district criterion tells us that, given that $p(Y(a)=y)$ is identifiable,
identification of $p(Y(a,M(a'))=y)$ may be obtained if the instruments are in separate districts and if neither of the instruments affects the other.
The associated cross-world assumption $Z(a)\ci L(a')$ indeed formalizes the need for no unmeasured confounding between the two instruments.
However, this cannot be guaranteed unless both $L$ and $Z$ are deterministic functions of (a randomized) treatment \cite{Robins2010}.
In that case, both $Z(a)$ and $L(a')$ are constants, and hence trivially independent.\footnote{In addition, as shown in \cite{Robins2010}, independence of $Z(a)$ and $L(a')$ leads to cross-world assumption~\eqref{ccwass}.}

Ironically, this required determinism seems to leave us incapable of pulling apart the pathways that we meant to separate in the first place.
However, progress can be made if one can conceive of separate interventions on $L$ and $Z$ that would enable to break their perfect correlation.
From this perspective, the deterministic characterization of an expanded causal DAG, such as Figure~\ref{exampleDAG3}D,
gives rise to a specific type of experimental design that requires one to think of $L$ and $Z$ as inherent but distinct properties of the treatment,
which may be intervened on separately but, when combined, fully capture all of its active ingredients.
The feasibility of such designs primarily mirrors the extent to which different active components of treatment can be conceived of being manipulated in isolation \citep{Didelez2013}.
Moreover, when combined with the aforementioned exclusion restrictions, such designs thus entail separate manipulations of $L$ and $Z$,
which capture distinct but exhaustive features of treatment to which, respectively, solely $M$ or $Y$ are (directly) responsive.
Importantly, this characterization enables to interpret natural effects as specific interventional contrasts.

Consider the causal DAG $\mathcal{G}(\mathbf{V}\cup \mathbf{H})$ with observed variables $\mathbf{V} = \{A,M,Y\}$ and hidden variable $\mathbf{H} = \{U\}$ in Figure~\ref{exampleDAG1}A,
and its deterministic expansion $\mathcal{G}'(\mathbf{V'}\cup \mathbf{H})$, with $\mathbf{V'} = \mathbf{V}\cup \{Z,L\}$, in Figure~\ref{exampleDAG3}D.
Let $Z$ and $L$ be deterministic functions which can be conceived as two complementary components that fully characterize what we will refer to as the `composite' treatment $\mathbf{A}$
such that $\mathbf{A} \equiv \{L,Z\}$, $\mathbf{a} \equiv \{a_L,a_Z\}$, $\mathbf{a'} \equiv \{a'_L,a'_Z\}$, $p(l\vert \mathbf{\tilde{a}}) = \mathbf{1}\{l=\tilde{a}_L\}$ and $p(z\vert \mathbf{\tilde{a}}) = \mathbf{1}\{z=\tilde{a}_Z\}$.
As pointed out by Robins and Richardson \cite{Robins2010}, $p(Y(a,M(a'))=y)$ then corresponds to the interventional distribution $p(Y(a_Z,a'_L)=y)$ since
\begin{align}
	p(Y(a,M(a'))=y) &= \sum_{z,l,m} p(y,m\vert \text{do}(z,l)) p(z\vert \text{do}(\mathbf{a})) p(l\vert \text{do}(\mathbf{a'}))\nonumber\\
	&= \sum_{z,l,m} p(y\vert z,l,m) p(m\vert l) p(z\vert \mathbf{a}) p(l\vert \mathbf{a'}) \label{jointintfunctional}\\
	&= \sum_{z,l} p(y\vert z,l) \mathbf{1}\{z=a_Z\} \mathbf{1}\{l=a'_L\} \nonumber\\
	&= p(y\vert a_Z,a'_L) = p(Y(a_Z,a'_L)=y).
\end{align}
The first equality is obtained by application of Equation~\ref{edgegform2}, the third by conditional independence $M\ci Z\vert L$ and determinism, and the second and last by Tian's \textbf{ID} algorithm.
Note that, because $Z(a) \ci L(a')$ holds by determinism rather than by independence restrictions implied by NPSEMs,
this result can be obtained under the `single world' model associated with the causal DAG in Figure~\ref{exampleDAG3}D.

The above result implies that, if deterministic mediating instruments like $L$ and $Z$ can be assumed to exist and the aforementioned exclusion restrictions are deemed plausible,
it is not necessary to actually conduct any experiment, nor to assume any cross-world independencies to identify the interventional distribution $p(Y(a_Z,a'_L)=y)$.
Instead, if $Y(a_Z,a'_L) = Y(a,M(a'))$ under $\mathcal{G}'(\mathbf{V'}\cup \mathbf{H})$ with $\mathbf{V'} = \mathbf{V}\cup \{Z,L\}$,
then identification of $p(Y(a_Z,a'_L)=y)$ is tantamount to identification of $p(Y(a,M(a'))=y)$ from observed data on $\mathbf{V}$
under the `single world' causal model representation of $\mathcal{G}(\mathbf{V}\cup \mathbf{H})$ \citep{Robins2010}.
For instance, if one merely assumes the existence of some (unidentified) deterministic mediating instruments $L$ and $Z$ in Figure~\ref{exampleDAG3}D,
measurements on $L$ and $Z$ are typically missing and $p(a'_L,a_Z) = 0$ (for $a'_L \ne a_Z$) in the observed sample.
Consequently, we can only express Equation~\ref{jointintfunctional} in terms of observable data on $\mathbf{V} = \{A,M,Y\}$ in the absence of unmeasured mediator-outcome confounding by $U$.
To recover identifiability we will thus generally need to complement $\mathbf{V}$ with an additional set of observable auxiliary variables of the types described in section~\ref{auxvar}.

\subsection{Some examples}
Some existing designs, such as double-blind placebo-controlled trials,
were in fact devised in the spirit of Robins and Richardson's \cite{Robins2010} deterministic extended graphs \citep{Didelez2013}.
Such trials aim to isolate part of the effect of the drug $A$ that may be attributed to active chemical components $Z$, and is not mediated by the patient's or doctor's expectations about the effectiveness of the drug $M$.
In such designs it is often reasonable to assume that expectations are solely affected by the knowledge of (possibly)
being treated $L$ and that the active component itself does not affect expectations.
The natural direct effect of the drug, not mediated by expectations, could therefore be interpreted as the interventional contrast comparing drug effectiveness between the treatment and placebo arm.
Note that experimental designs that reflect an expanded deterministic graph do not require any measurements on the mediator to identify interventional contrasts that correspond to certain natural effects.

Unfortunately, success is not always guaranteed.
Side effects in the treatment arm may, for instance, raise suspicions of being on active treatment, thereby violating
the crucial exclusion restriction $M\ci Z\vert L$ encoded in Figure~\ref{exampleDAG3}D.
To accommodate for known side effects, active placebos have been designed that mimick side effects of the active treatment \citep{didelezsymposium,Lok2016},
illustrating that the ability to increase the credibility of required exclusion restrictions may often be highly dependent on the creativity of the researcher \citep{Robins2010}.\footnote{In a strict sense, active placebo designs also violate the required exclusion restrictions.
Nonetheless, they enable to arrive at a measure of a direct effect that more closely resembles the natural direct effect of primary interest \citep{didelezsymposium}.}

In other contexts, experimental designs in the spirit of deterministic extended graphs are more difficult to conceive.
For instance, even though the JOBS II study \citep{Vinokur1997} involved a job search skills workshop that targeted specific component processes
grounded in psychological theory, it may still be hard to imagine similar interventions or workshops that isolate the distinct triggering elements of separate targeted processes,
let alone, to conceive of distinct elements that exclusively affect either re-employment or mental health (via direct pathways).
Any attempt to endow natural direct and indirect effects with an interventionist interpretation
would thus necessarily rely on strong theoretical assertions about the active components of the job training intervention.

\section{Path-specific effects for multiple mediators \label{pathspeceff}}
The focus of this chapter has hitherto been restricted to identification of natural effects.
The recanting district criterion, however, delineates conditions that permit identification of
any effect along any bundle of pathways that may be of interest.
Its utility may thus be particularly appealing in settings with intertwined pathways along multiple mediators or longitudinal settings where both treatment and/or mediators may be time-varying.

Because of the inherent cross-world nature of path-specific effects,
non-parametric identification necessarily always relies on untestable cross-world independence assumptions.
A major appeal of the recanting district criterion is that it makes explicit formulations
of relevant path-specific cross-world independence assumptions essentially redundant for the purpose of identification
under NPSEM representations of hidden variable DAGs, as illustrated below.
In general, decompositions of the treatment effect into path-specific effects other than natural direct and indirect effects may be motivated by non-identifiability
of natural effects or by the simple fact that the primary mediation hypothesis cannot be expressed in terms of such effects.

\paragraph{Alternative decompositions in the presence of intermediate confounding}
In our motivating example, the natural indirect effect with respect to re-employment $M$ is not identified if re-employment and mental health $Y$
are believed to be subject to treatment-induced confounding by changes in perceived self-efficacy, as denoted by $L$ in Figure~\ref{exampleDAG2}A.
If, nonetheless, the main interest is in the mediating role of re-employment, we may either calculate partial identification bounds for the natural indirect effect (see \citep{Miles2017} and references therein),
conduct a sensitivity analysis (see \citep{VanderWeele2014a} and references therein), or abandon focus on the natural indirect effect altogether.
Instead, shifting focus to less ambitious decompositions in terms of either the joint natural indirect effect mediated by both perceived self-efficacy and re-employment
(transmitted by $\pi_1 = \{A\rightarrow M\rightarrow Y; A\rightarrow L\rightarrow Y; A\rightarrow L\rightarrow M\rightarrow Y\}$),
or the partial indirect effect with respect to re-employment (transmitted by $\pi_2 = \{A\rightarrow M\rightarrow Y\}$),
may still to some extent enable us to learn about the mediating role of re-employment.

The \emph{joint natural indirect effect} with respect to $\{L,M\}$ requires recovering the cross-world counterfactual distribution $p(Y(\pi_1,a,a')=y) = p(Y(a',L(a),M(a))=y)$ and
is identifiable even if we relax assumptions to allow for unmeasured confounding between sense of self-efficacy and re-employment, as in Figure 3.4C.
This can easily be seen upon noting that the district $\{L,M\}$ is not recanting with respect to $\pi_1$.
In this case, the recanting district criterion serves to establish cross-world independence $$Y(a,l,m) \ci \{M(a',l), L(a')\}.$$
When combined with experimentally verifiable identifying assumptions for $p(Y(a)=y)$, as encoded in the `single-world' representation of the hidden variable DAG,
this cross-world assumption renders $p(Y(a',L(a),M(a))=y)$ identifiable from observed data.
In contrast, districts $\{M,Y\}$ and $\{L,Y\}$ are recanting with respect to $\pi_1$, as in Figures 3.4B and D respectively,
such that $p(Y(a',L(a),M(a))=y)$ is not identifiable under unmeasured confounding
of the relation between the outcome and any of the given intermediate variables along paths in $\pi_1$.
Nonetheless, identification can be restored in the presence of a mediating instrument, for instance, on the edge $A\rightarrow M$ or $A\rightarrow Y$ if it is hindered due to unmeasured $M-Y$ confounding.

The \emph{partial indirect effect} with respect to re-employment, on the other hand, requires recovering $p(Y(\pi_2,a,a')=y) = p(Y(a',L(a'),M(a,L(a')))=y)$.
This cross-world counterfactual distribution remains identifiable if we relax assumptions by allowing for unmeasured confounding between
sense of self-efficacy and mental health, as in Figure 3.4D, because the district $\{L,Y\}$ is not recanting with respect to $\pi_2$.
Allowing for unmeasured confounding of the relation between re-employment $M$ and either the outcome $Y$ or intermediate confounder $L$,
in contrast, destroys identification because of recantation of the districts $\{M,Y\}$ and $\{L,M\}$ with respect to $\pi_2$, in Figures 3.4B and C, respectively.
The violated cross-world independence assumption $$\{Y(a,l,m),L(a)\} \ci M(a',l)$$
which enables identification of $p(Y(a',L(a'),M(a,L(a'))=y)$, can, however, again be interchanged with another assumption which restores identification in the presence of a mediating instrument on one of the respective edges emanating from $A$.

\paragraph{Addressing different types of mediation questions}
In certain cases, the partial indirect effect with respect to a mediator of interest $M$ that is affected by earlier mediators, may, however, be the primary path-specific effect of interest.
For example, Miles and colleagues \citep{Miles2017a} aimed to assess the extent to which treatment adherence driven by non-toxicity factors mediates the effect of antiretroviral therapy (ART) on virological failure in HIV patients in Nigeria.
Ignoring potential baseline confounders, their target of inference corresponds to the path-specific effect along $\pi = \{A\rightarrow M\rightarrow Y\}$ in Figure 3.4D, where $A$ denotes ART, $L$ drug toxicity, $M$ adherence and $Y$ viral load.
Their corresponding mediation analysis thus aimed to answer the question \textit{``How much of the medication's effect is mediated by adherence, if we discard the mediating role of adherence driven by drug toxicity?''}
Estimation of this contribution to the total effect of ART on viral load enabled Miles and colleagues to address questions that are not only etiologically relevant but that may also have important policy implications.
In fact, the corresponding mediation analysis aimed to assess whether conceivable modifications to the treatment regimen, which may increase adherence (but not through changes in toxicity),
may magnify the net treatment effect and hence increase its effectiveness.
The interpretation of the partial indirect effect as an interventional contrast, which could be estimated from such a hypothetical experiment, can likewise be represented via a deterministic expanded DAG.
A more detailed discussion of deterministic expanded DAGs for path-specific effects, as discussed in \citep{Robins2010}, is, however, beyond the scope of this chapter.

\section{Discussion and further challenges}
Most developments on the identification of natural direct and indirect effects have so far focused on single mediator settings where cross-world independence~\eqref{ccwass} holds
along with conditional ignorability assumptions~\eqref{adjform1} and~\eqref{adjform2}.
That is, where the data-generating mechanism can be described by an NPSEM in which a common set of baseline covariates $\mathbf{C}$ suffices to adjust for confounding of the
treatment-mediator, treatment-outcome and mediator-outcome associations (within levels of treatment),
and where moreover none of the elements of $\mathbf{C}$ is affected by treatment.
The latter two requirements have, to a large extent, prohibited extensions to settings with multiple, possibly longitudinal, mediators.

Recently, a complete graphical criterion has been devised for identification of any path-specific effect of a treatment on an outcome under NPSEMs \citep{Shpitser2013}.
Briefly, it shows that when the total causal effect is identifiable by some means, as can be verified using Tian's \textbf{ID} algorithm \citep{Tian2003}
(as discussed in the previous chapter), then every path-specific effect (along a set of pathways) for which there is no recanting district is also identifiable.
Identification then essentially proceeds via Tian's identifying functional (Equation~\ref{gform2}), which extends Robins' g-functional \citep{Robins1986} to (NPSEM representations of) hidden variable causal DAGs,
while allowing treatment assignments to be different across districts (Equation~\ref{edgegform2}).

\paragraph{Increased identification power}
It is not too hard to come up with examples of settings where the adjustment criterion for natural effects fails to identify $p(Y(a,M(a'))=y)$,
due to violations of assumption set~\eqref{adjform1}-\eqref{adjform2},
yet the recanting district criterion leads to identifiable natural direct and indirect effects (see Figures~\ref{Pearl2014fig5fsingle} and~\ref{exampleDAG3}, respectively).
Even so, from a practitioner's point of view, it can be argued that the associated increased identification power is of limited practical relevance in single mediator settings.
The reason is that prior knowledge in practice is often too limited to justify assumptions that could substitute for failure of assumptions~\eqref{adjform1}-\eqref{adjform2} \citep{Imai2014}.
On the other hand, one cannot ignore the potential of causal structure learning algorithms (see chapter 4), especially in the `big data' age, in which we have increasing access to massive data sets.
In particular, such algorithms may aid researchers to construct a class of DAGs that are compatible with the observed data distribution and
under which identification may be obtained under more general assumptions as delineated by the \textbf{ID} algorithm and the recanting district criterion.

Importantly, the recanting district criterion has been proposed as a criterion that, given identifiability of $p(Y(a)=y)$, delineates conditions for identifying marginal (or population-averaged) path-specific distributions $p(Y(\pi,a,a')=y)$.
Nonetheless, its utility is still unclear (but definitely more subtle) when it comes to identification of conditional (or stratum-specific) path-specific distributions $p(Y(\pi,a,a')=y\vert \mathbf{c})$.
Generalizations for complete identification criteria for conditional path-specific effects are undoubtedly less straightforward, and are left as subject for further research.
Consequently, the increased identification power that follows from Shpitser's results is currently only well-documented for marginal natural direct and indirect effects,
since the adjustment criterion for natural direct and indirect effects delineates identical conditions for identifying both $p(Y(a,M(a'))=y)$ and $p(Y(a,M(a'))=y\vert \mathbf{c^*})$
whenever $\mathbf{C^*}$ is a subset of a set of baseline covariates $\mathbf{C}$ that controls for mediator-outcome confounding.

\paragraph{Estimation}
If identification of natural effects is achieved under the above `traditional' but more stringent set of assumptions, this leads to a standard identifying functional,
generally known as the mediation formula, for which a well-established suite of (semi-)parametric estimators has been developed (see \citep{Vansteelandt2012f} and references therein).
Accordingly, estimation may then proceed via routine application of these methods as implemented in off-the-shelf statistical software packages (see \citep{Steen2017} and references therein).
Even though software implementations of complete graphical identification algorithms, such as the \textbf{ID} algorithm, are now publicly available \citep{Tikka2017},
estimation of more involved or less standard identifying functionals arguably imposes another barrier to routine application of complete identification algorithms.
Inevitably, this may have led to a trade-off between postulating realistic causal structure, on the one hand, and simple and accessible estimation strategies, on the other hand.
Future research thus needs to focus on the development of a generic and flexible estimation framework for more generic identifying functionals.
Such framework should not only incorporate estimation of natural effects, but also of more generally defined path-specific effects
(see \citep{Miles2017a,Steen2017a} for some first promising steps in this direction).

\paragraph{Broadening the scope}
Even though its added value for identification of natural effects may be debatable,
the recanting district criterion offers a major potential for extensions to settings with multiple, possibly longitudinal, mediators.
The identification of the partial indirect effect via a given mediator of interest, as in Figure~\ref{exampleDAG2}D, forms a first step towards such extension,
as it allows for possibly high-dimensional post-treatment confounders to confound the mediator-outcome association.
In particular, it allows for earlier mediators to be confounders of the association between later mediators and outcome,
while at the same time being confounded with the outcome by unmeasured common causes.

\paragraph{Cross-world contemplations}
Both the cross-world nature of path-specific effects and the required cross-world independence assumptions for identification
have been the subject of an ongoing debate \citep[e.g.][]{Naimi2014c,Robins2010}, roughly dividing the field into NPSEM `skeptics' and `advocates'.
We have tried to shed some light on this controversy, and illustrated the important role of mediating instruments and deterministic expanded graphs \citep{Robins2010}
in elucidating and bridging this conceptual and ontological divide.
The main objection is that such cross-world independence assumptions, on which modern causal mediation analysis generally relies (although see \citep{TchetgenTchetgen2017} for a recent exception),
cannot be enforced experimentally and hence are not falsifiable.
However, whether or not researchers should be encouraged to reformulate their mediational hypotheses in terms of feasible potential interventions on defining features of treatment, remains an open question.

An alternative approach to avoiding cross-world definitions and assumptions has recently gained increasing attention.
This approach builds on the claim that, even in the absence of any reference to cross-world quantities or restrictions,
certain contrasts based on the mediation formula may still carry empirically meaningful interpretations \citep{Didelez2006,Petersen2006}.
This has given rise to the more formal definition of so-called \emph{randomized intervention analogs} of natural effects \citep{VanderWeele2014,VanderWeele2016,Vansteelandt2017},
which conceive of setting the mediator at some level that is randomly assigned from the conditional counterfactual mediator distribution $p(M(a')=m\vert \mathbf{c})$ rather than at the individual counterfactual level (see \citep{Lok2016} for a related approach).
Importantly, because their definitions do not employ cross-world counterfactuals strong and unfalsifiable assumptions,
such as cross-world independence (but also `no intermediate confounding') may be avoided.
These estimands also tend to correspond more closely to relevant policy measures that can be estimated from actual interventions.

Even if one is willing to make untestable cross-world assumptions,
identification of natural effects is typically hindered in the presence of treatment-induced confounding.
Accordingly, several contributions to the field have articulated alternative assumptions that may
allow us to recover natural effects despite treatment-induced confounding.
However, such assumptions generally impose additional structure on the joint distribution of counterfactuals,
such as rank preservation \citep{Robins2010}, monotonicity \citep{TchetgenTchetgen2014a} or parametric constraints \citep{Greenland1992,Petersen2006,TchetgenTchetgen2014a}.

\section{Acknowledgements}
The authors would like to thank Vanessa Didelez, Yves Rosseel, Karel Vermeulen and one anonymous referee for helpful suggestions and feedback on an earlier draft,
and Ilya Shpitser for valuable discussions that have led to some improved insights presented in this chapter.
This work was supported by Research Foundation Flanders (FWO Grant G.0111.12).

\bibliographystyle{plain}
\bibliography{ms}

\end{document}